\documentclass[runningheads]{llncs}

\usepackage[T1]{fontenc}
\usepackage{graphicx}
\usepackage{verbatim}
\usepackage{placeins}
\usepackage{lipsum}
\usepackage{hyperref}
\hypersetup{
    colorlinks=true,
    linkcolor=blue,
    filecolor=magenta,      
    urlcolor=cyan,
    pdftitle={Overleaf Example},
    pdfpagemode=FullScreen,
    }

\usepackage{chngcntr}
\usepackage{ulem}
\usepackage{booktabs}
\usepackage{multirow}
\usepackage{tikz}
\usepackage{amsmath}
\usepackage{amssymb}
\usepackage{tablefootnote}

\def\checkmark{\tikz\fill[scale=0.4](0,.35) -- (.25,0) -- (1,.7) -- (.25,.15) -- cycle;}

\newcommand{\cs}[1]{\texttt{#1}}

\begin{document}

\title{GEPAR3D: Geometry Prior-Assisted Learning for 3D Tooth Segmentation}
%
\begin{comment}  %% Removed for anonymized MICCAI 2025 submission
\author{First Author\inst{1}\orcidID{0000-1111-2222-3333} \and
Second Author\inst{2,3}\orcidID{1111-2222-3333-4444} \and
Third Author\inst{3}\orcidID{2222--3333-4444-5555}}
%
\authorrunning{F. Author et al.}
% First names are abbreviated in the running head.
% If there are more than two authors, 'et al.' is used.
%
\institute{Princeton University, Princeton NJ 08544, USA \and
Springer Heidelberg, Tiergartenstr. 17, 69121 Heidelberg, Germany
\email{lncs@springer.com}\\
\url{http://www.springer.com/gp/computer-science/lncs} \and
ABC Institute, Rupert-Karls-University Heidelberg, Heidelberg, Germany\\
\email{\{abc,lncs\}@uni-heidelberg.de}}

\end{comment}

\author{Tomasz Szczepański\inst{1} \and
Szymon Płotka \inst{1,2} \and
Michal K. Grzeszczyk \inst{1} \and
Arleta Adamowicz \inst{3} \and
Piotr Fudalej \inst{3} \and
Przemysław Korzeniowski \inst{1} \and
Tomasz Trzciński \inst{4,5} \and
Arkadiusz Sitek \inst{6}
}
\authorrunning{T.Szczepański et al.}

\institute{Sano Centre for Computational Medicine, Cracow, Poland \\
\email{t.szczepanski@sanoscience.org} \and
Jagiellonian University, Cracow, Poland \and
Jagiellonian University Medical College, Cracow, Poland \and
Warsaw University of Technology, Warsaw, Poland \and
Research Institute IDEAS, Warsaw, Poland \and
Massachusetts General Hospital, Harvard Medical School, Boston, MA, USA \\
}

\maketitle  
\begin{abstract}
Tooth segmentation in Cone-Beam Computed Tomography (CBCT) remains challenging, especially for fine structures like root apices, which is critical for assessing root resorption in orthodontics. We introduce \texttt{GEPAR3D}, a novel approach that unifies instance detection and multi-class segmentation into a single step tailored to improve root segmentation. Our method integrates a Statistical Shape Model of dentition as a geometric prior, capturing anatomical context and morphological consistency without enforcing restrictive adjacency constraints. We leverage a deep watershed method, modeling each tooth as a continuous 3D energy basin encoding voxel distances to boundaries. This instance-aware representation ensures accurate segmentation of narrow, complex root apices. Trained on publicly available CBCT scans from a single center, our method is evaluated on external test sets from two in-house and two public medical centers. \texttt{GEPAR3D} achieves the highest overall segmentation performance, averaging a Dice Similarity Coefficient (DSC) of 95.0\% (+2.8\% over the second-best method) and increasing recall to 95.2\% (+9.5\%) across all test sets. Qualitative analyses demonstrated substantial improvements in root segmentation quality, indicating significant potential for more accurate root resorption assessment and enhanced clinical decision-making in orthodontics. We provide the implementation and dataset at \href{https://github.com/tomek1911/GEPAR3D}{\texttt{github.com/tomek1911/GEPAR3D}}.
\end{abstract}
\section{Introduction}
CBCT is essential in digital dentistry, yet manual tooth segmentation remains labor-intensive and inconsistent~\cite{zheng2024semi}. Automated methods support treatment planning and diagnostics~\cite{kapila2011current}, but delineating tooth roots remains challenging due to their intricate morphology and small size. Accurate segmentation is particularly important for assessing root resorption~\cite{samandara2019evaluation}, a pathological loss of dentin and cementum often caused by orthodontic tooth movement, which can weaken stability and, in severe cases, increase the risk of tooth loss. Tooth geometry exhibits universal patterns, with teeth arranged in two arches and four quadrants, and although individual variability exists, key anatomical features remain stable~\cite{kimura2009common}. Demonstrating this structural consistency, the upper molars typically have three roots, while the lower ones usually have two~\cite{vertucci1984root}. Leveraging inherent geometric priors may provide valuable guidance for enhancing the accuracy and robustness of automated segmentation.\\
\begin{figure}[t!]
\centering
\includegraphics[width=\textwidth]{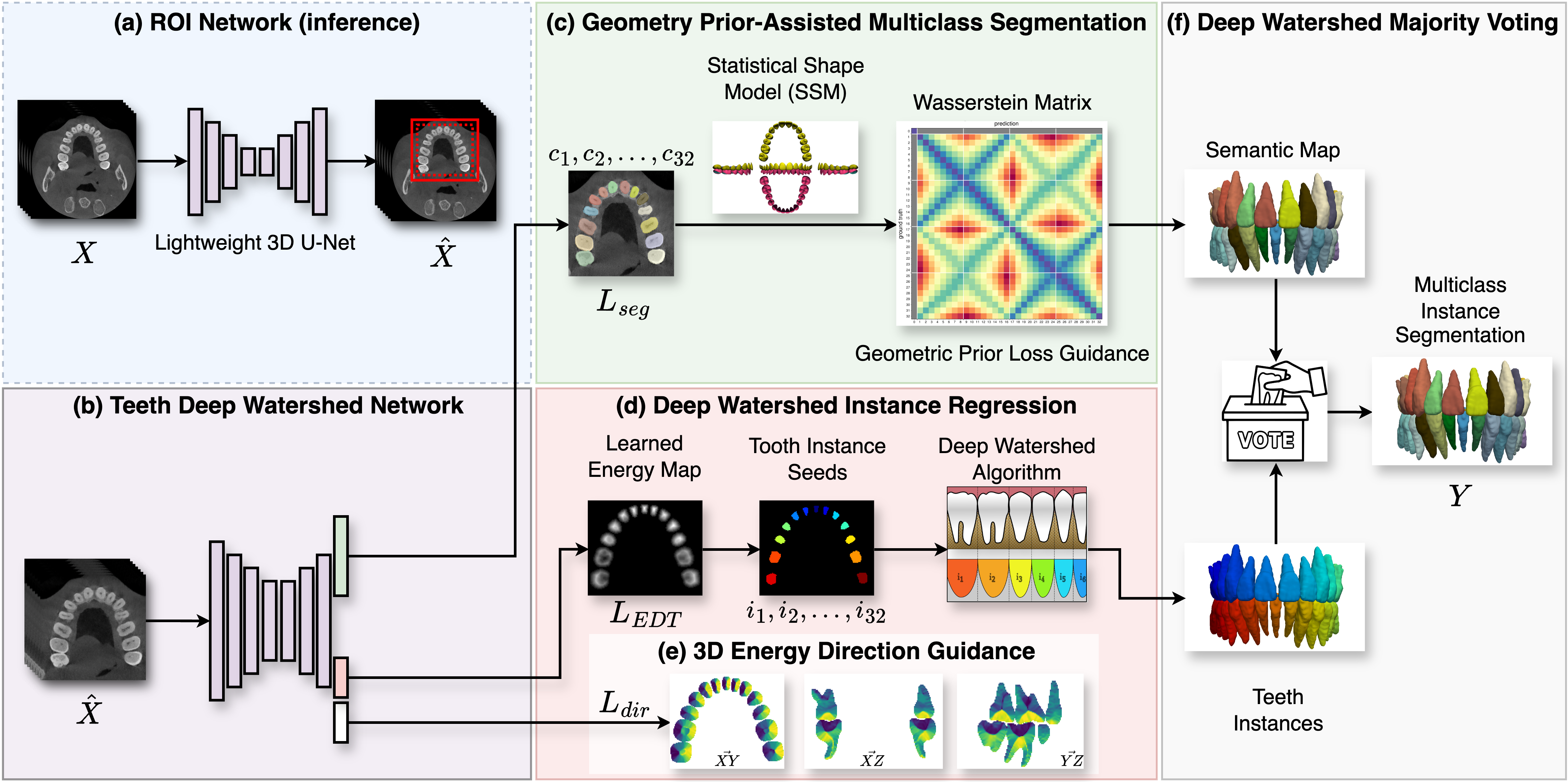}
\caption{An overview of \texttt{GEPAR3D}, which unifies instance detection and multi-class segmentation for precise tooth root segmentation. \textbf{(a)} Crops the region of interest (ROI) during inference. \textbf{(b)} Simultaneously performs multi-class segmentation and instance regression. \textbf{(c)} Regularizes segmentation loss $L_{seg}$ with a geometric prior from an SSM of normal dentition~\cite{kim2022developing}. \textbf{(d)} Uses instance regression task $L_{EDT}$ to generate energy maps for the Deep Watershed Algorithm. \textbf{(e)} Captures complex root apex geometries via Energy Direction loss $L_{dir}$. Finally, \textbf{(f)} assigns each detected instance a class via majority voting based on segmentation outputs.}
\label{fig:overview}
\end{figure}%
\indent
Tooth segmentation has evolved from heuristic, hand-crafted methods to deep learning~\cite{polizzi2023tooth}. Early deep learning approaches rely on voxel-wise classification and overlook anatomical structure and inter-class relationships, both essential for capturing detailed tooth morphology, especially root regions~\cite{cui2019tooth,lee2020automated,chen2020automatic,szczepanski2024let}. To reduce computational costs, many methods use multi-step coarse-to-fine pipelines that isolate individual teeth in bounding boxes~\cite{chung2020pose,cui2021hierarchical,jang2021fully,kim2023automatic,tan2024progressive}. Such pipelines tend to accumulate errors and disconnect each tooth from its broader anatomical context, impairing root apex segmentation. Recent methods integrate prior anatomical knowledge to guide segmentation. ToothSeg~\cite{cui2021hierarchical} enforces shape consistency via tooth skeletons but relies on manual thresholds and post-processing, limiting generalization. Other methods incorporate spatial relationships via adjacency constraints; SGANET~\cite{li2022semantic} uses rigid graph convolutions to enforce local consistency of immediate neighbors, while TSG-GCN~\cite{liu2024individual} learns dynamic adjacency but remains vulnerable to data bias since its geometric priors derive solely from training labels. These constraints highlight the need for a method that integrates anatomical structure and inter-class dependencies to achieve a coherent, context-rich representation for accurate root segmentation.\\
\indent
To achieve that, we propose \textbf{GE}ometric \textbf{P}rior-\textbf{A}ssisted Lea\textbf{R}ning for \textbf{3D} (\texttt{GEPAR3D}), a novel approach that unifies instance detection and multi-class segmentation in a single step. Our method integrates a Statistical Shape Model (SSM) \cite{kim2022developing} of dentition as a geometric prior, based on inter-teeth distances, to embed anatomical context and morphological consistency into the learning process. Furthermore, we leverage a deep watershed method and model each tooth as a continuous 3D energy basin, encoding voxel distances to boundaries, and predicting directional gradients to capture subtle variations at root apices.
Trained on publicly available CBCT scans from a single center and evaluated on external test sets from four medical centers (two in-house and two public), \texttt{GEPAR3D} demonstrates robust generalization across diverse patient demographics. Our method outperforms five state-of-the-art methods, achieving the highest segmentation performance with an average DSC of 95.0\% (+2.8 \%) and RC of 95.2\% (+9.5 \%) across all test sets, offering new prospects for reliable root resorption assessment and improved clinical decision-making.

\section{Methodology}
Fig. \ref{fig:overview} overviews \texttt{GEPAR3D}, an encoder-decoder model with dual decoders for multi-class segmentation and instance regression. The segmentation branch classifies 32 tooth categories with SSM-based regularization, while the regression branch models instances as energy basins guided by energy descent. Each detected instance receives class votes from the multi-class segmentation branch, and the final class assignment is determined through majority voting.\\
\noindent
\textbf{Geometric prior.} To enhance root segmentation in CBCT scans, we integrate a Statistical Shape Model (SSM) of normal dentition~\cite{kim2022developing} as a geometric prior. Built from a representative patient cohort, this 3D atlas encodes anatomical knowledge. By processing the SSM, we extract inter-tooth distances for morphologically guided segmentation.
To capture statistical tooth positions, we represent each tooth’s geometric center as $G_i = (x_i, y_i)$ in a normalized coordinate system, where $i$ denotes the tooth index. Each quadrant $Q_k$, where $k \in \{1, 2, 3, 4\}$, contains 8 teeth, defined as $T_{Q_k} = \{G_{k1}, G_{k2}, \dots, G_{k8}\}$. The statistical inter-tooth Euclidean distances $D_{ij}^{(k)}$ within a quadrant are computed as: $D_{ij}^{(k)} = \sqrt{(x_i - x_j)^2 + (y_i - y_j)^2}, \quad \text{with } G_i, G_j \in T_{Q_k}.$ The origin $O_{xy}$ of the normalized system is set at the midpoint of the maxillary and mandibular central incisors, whose geometric centers define: $O = \frac{1}{4} \left( G_{11} + G_{21} + G_{31} + G_{41} \right).$ Since $T_{k,8}$ is absent from the SSM due to rarity, its geometric center $G_8$ is interpolated from $G_6$ and $G_7$. Finally, statistical inter-tooth distances \(D_{ij}\) for each quadrant \(Q_k\) are obtained by averaging male and female dentition models, forming the intra-quadrant distance matrix \(\mathbf{D}_{Q_k} = [D_{ij}^{(k)}]\), where \(D_{ij}\) encodes pairwise Euclidean distances between teeth within \(Q_k\).\\
\noindent
\textbf{Geometry Prior-Assisted Learning.} 
To ensure semantically meaningful predictions, we adapt a prior-regularized Dice Similarity Coefficient (DSC) for multi-class segmentation, the Generalized Wasserstein Dice Loss (GWDL)~\cite{fidon2018generalised}, by replacing its original empirical dissimilarity penalties with statistical inter-tooth distances as a geometric prior, yielding the Geometric Wasserstein Dice Loss (GeoWDL). Derived from Optimal Transport (OT) theory, Wasserstein Distance (WD) measures the minimal cost of transforming one probability distribution into another, and now enriched with geometrical prior allows for a structured penalization of segmentation errors based on spatial and morphological relationships. It penalizes misclassifications based on spatial and morphological relationships: higher penalties are assigned to errors between distant teeth within the same quadrant, reflecting geometric prior, while semantically weighted adjustments assign lower penalties to misclassifications among morphologically similar teeth within the same arch and higher penalties to confusions between the structurally distinct upper and lower arches. We introduce penalty modifiers $p_{Q_{k_i}Q_{k_j}}$ to weight geometry-based penalties, as matrix $\mathbf{Q}_{q_{ij}}$ $(i,j \in k)$ to penalize confusions between quadrants $Q_k$: within dental arch ($p_{Q_1Q_2}=0.1$), between arches ($p_{Q_1Q_4}=0.2$) and diagonally ($p_{Q_1Q_3}=0.3$), given as:

\begin{equation}
\label{eq:quadratn_penalty_matrix}
\begin{aligned}
& \mathbf{Q}_{q_{ij}}= \left[\begin{array}{cccc}
0 & q_{12} & q_{13} & q_{14} \\
q_{21} & 0 & q_{23} & q_{24} \\
q_{31} & q_{32} & 0 & q_{34} \\
q_{41} & q_{42} & q_{43} & 0
\end{array}\right] = \left[\begin{array}{cccc}
0 & 0.1 & 0.3 & 0.2 \\
0.1 & 0 & 0.2 & 0.3 \\
0.3 & 0.2 & 0 & 0.1 \\
0.2 & 0.3 & 0.1 & 0
\end{array}\right].
\end{aligned}
\end{equation}
\noindent
To obtain the penalty matrix $\mathbf{M_{\text{geo}}}$ (hereafter $M$ for brevity), we arrange $\mathbf{D}_{Q_k}$ and apply penalty modifiers $\mathbf{Q}_{q_{ij}}$. First, we define the helper matrix $\mathbf{P}_{mn,ij} = q_{ij} {J}_{mn}$, where \( {\mathbf{J}}_{mn} \) is an \( 8 \times 8 \) matrix of ones and \(m,n\) index index the elements of \(J\), ensuring that multiplication with \( q_{ij} \) results in a matrix filled with the corresponding penalty value. Next, we compute $\mathbf{M}_{k \times mn} = \mathbf{D}_{Q_k}$ + $\mathbf{P}_{mn}$. The resulting $\mathbf{M}$ is $33\times33$, with 32 tooth classes ($l$) normalized to (0,1). A background class ($b$) is added, with $b=2$ to strongly penalize tooth-to-background miss-classification, followed by final normalization for consistency.
We integrate the geometrical and morphological prior of $M$ within the loss function as follows:
\begin{equation}
\begin{aligned}
&L_{GeoWDL}(\hat{\mathbf{p}}, \mathbf{p})  = 1-\frac{2 \sum_{l} \sum_i \mathbf{p}_{i, l}\left(1-W^M\left(\hat{\mathbf{p}}_i, \mathbf{p}_i\right)\right)}{2 \sum_{l}
\left[\sum_i p_{i, l}\left(1-W^M\left(\hat{\mathbf{p}}_i, \mathbf{p}_i\right)\right)\right]+\sum_i W^M\left(\hat{\mathbf{p}}_i, \mathbf{p}_i\right)},
\end{aligned}
\label{eq:gwd}
\end{equation} 
\noindent where $W^{M}(\hat{p}_i, p_i)$, given as: $W^M\left(\hat{\mathbf{p}}_i, \mathbf{p}_i\right) = \sum_{l=1}^L p_{i,l} \sum_{l^{\prime}=1}^L M_{l, l^{\prime}} \hat{p}_{i, l^{\prime}}$ is the WD-weighted probability mass between predicted $\hat{p}_i$ and ground truth $p_i$ at voxel $i$. To address class imbalance in third molars, the final segmentation loss \( L_{\text{seg}} \) combines \( L_{\text{GeoWDL}} \) with inverse class frequency weighted cross-entropy \( L_{\text{WCE}} \).\\
\noindent
\textbf{Deep Watershed Instance Regression.}
To generate inputs for the deep watershed algorithm, our method optimizes two complementary tasks: energy basin regression and directional gradient estimation for boundary refinement. These tasks ensure accurate instance separation while encoding tooth boundaries and instance identity, enabling full tooth spatial understanding. We adapt the 2D deep watershed approach~\cite{bai2017deep} to 3D tooth instance segmentation, modeling each tooth as an energy basin with a smooth energy gradient. First, we replace discrete energy-level classification with continuous energy map regression, which encodes spatial structure by assigning each voxel a distance to the nearest tooth boundary. This contrasts with the original approach and~\cite{cui2021hierarchical,tan2024progressive}, which classify discrete offsets from the tooth centroid within a multi-task approach. Second, we refine boundary localization by estimating energy descent directions at each voxel, crucial for capturing rapid gradient changes in root apices. Unlike sequential pretraining~\cite{bai2017deep}, we train direction estimation as a parallel auxiliary task for efficiency. These adaptations aim to enable end-to-end optimization, improving instance awareness and refining segmentation precision, particularly in fine root structures. We compute watershed energy basins from ground truth using the Euclidean Distance Transform, which encodes distances to the instance boundary: $L_{EDT} = \frac{1}{N} \sum_{x=1}^{X} \sum_{y=1}^{Y} \sum_{z=1}^{Z} [I(x, y, z) - \hat{I}(x, y, z)]^2$. For energy direction, let \( E(\mathbf{r}) \), where \( \mathbf{r} = (x, y, z) \), be the energy map defining a scalar field over a 3D voxel grid. The gradient \( G(r) \) at each voxel \( r \) is computed by convolving the scalar field \( E(r) \) with the 3D Sobel-Feldman operator \( K_d \) along each dimension \( d \): $G(r) = \sum_{d \in \{x, y, z\}} K_d \ast E(r)$. The gradient magnitude \( G(r) \) and the unit direction vector \( \vec{u}_v \) for each voxel \( v \) are given by $G(\mathbf{r}) = \|G(r)\|_2$ and $\vec{u}_v = \frac{G(r)}{G(r)},$ where \( \|\cdot\|_2 \) denotes the Euclidean norm. Maximum angular error \( \theta_v = \pi \) occurs when \( \vec{u}_v \) misidentifies instance's center, refining boundary localization. We optimize the 3-channel decoder's $D$ output using the mean squared error loss in the angular domain: $L_{dir} = \sum_{v\in P_{l}} || \cos^{-1} \langle \vec{u}_{p \:\text{GT}}, \vec{u}_{p \:\text{pred}} \rangle||^2$, where $P_{l}$ and $l \in \{1,2,...,32\}$ is the set of all voxels belonging to the tooth semantic class. We mask non-tooth areas to reduce complexity and accelerate convergence. We clip $cos^{-1}$ to $\left<-1,1\right>$, for numerical stability.\\
\noindent
\textbf{Deep Watershed Instance Classification via Majority Voting.}
Since instance segmentation assigns a single, consistent label to each tooth, we first apply the deep watershed algorithm to separate tooth instances and then classify them using majority voting based on voxel-wise predictions from the semantic segmentation branch. To this end, we extract instance seeds by empirically thresholding energy basins at half their depth ($\beta = 0.5$). Next, we binarize the multi-class segmentation to create a mask, restricting the watershed algorithm to tooth regions for improved computational efficiency. We then apply the watershed algorithm using predicted energy maps, seed points, and the segmentation mask, derived from an end-to-end optimized model, to separate 3D tooth instances. Finally, each instance is assigned a class using majority voting, selecting the most frequent voxel-wise prediction. For a given instance \( j \), let \( C_i \) be the class of voxel \( i \) in segmentation \( S \), and \( Iv_j \) its volume:  \( \text{C}_j = \text{argmax}_c \sum_{i \in V_j} \delta(C_i, c) \), where \( \delta(C_i, c) \) is 1 if \( C_i = c \), otherwise 0. The class with the highest count is assigned to the instance.
\section{Experiments and results}
\begin{table}[t!]\centering
\caption{
Quantitative results of \texttt{GEPAR3D} and state-of-the-art methods for general and tooth-specific segmentation. We report Detection Accuracy (DA) and F1 for instance detection and classification; DSC, RC, and HD for multi-class segmentation; and $NSD_1$ and $RC_B$ for binary segmentation, with means and standard deviations (in brackets). Results are averaged across three external datasets unless stated explicitly. Methods are sorted by average DSC; best and second-best tooth-specific methods are highlighted in bold and underlined, respectively. \textbf{I}, \textbf{S}, and \textbf{IS} denote instance, semantic, and instance-based multi-class segmentation. $\dagger$ indicates p-value $<$ 0.05.
}
\resizebox{\textwidth}{!}{%
\begin{tabular}{@{}lclcclcccclcclcc@{}}
\toprule
\multirow{2}{*}{Method} & \multirow{2}{*}{Type} &  & \multirow{2}{*}{DA (\%)$\uparrow$} & \multirow{2}{*}{F1 (\%)$\uparrow$} &  & \multicolumn{4}{c}{DSC (\%)$\uparrow$} &  & \multirow{2}{*}{RC (\%)$\uparrow$} & \multirow{2}{*}{HD (mm)$\downarrow$} &  & \multirow{2}{*}{$NSD_1$ (\%)$\uparrow$} & \multirow{2}{*}{$RC_B$ (\%)$\uparrow$} \\
\cmidrule(lr){7-10}
 &  &  &  &  &  & In-house & Cui et al. & TF2 & Average &  &  &  &  &  &  \\
\cmidrule(r){1-2} \cmidrule(lr){4-5} \cmidrule(lr){7-10} \cmidrule(l{0pt}r{2pt}){11-12} \cmidrule(l{2pt}r{3pt}){13-13} \cmidrule(l{3pt}r{2pt}){14-15} \cmidrule(l{2pt}r{0pt}){16-16}
U-Net $\dagger$ \cite{ronneberger2015u} & S &  & $95.6(4.6)$ & $93.5(6.3)$ &  & $87.8(3.4)$ & $88.7(3.3)$ & $88.3(3.1)$ & $88.2(3.8)$ &  & $85.5(5.0)$ & $21.78(11.24)$ & & $88.2(5.1)$ & $90.6(2.9)$ \\
Swin SMT $\dagger$\cite{plotka2024swin} & S &  & $98.1(3.1)$ & $96.8(4.0)$ &  & $92.8(2.5)$ & $92.9(2.5)$ & $91.4(3.0)$ & $92.3(2.8)$ &  & $91.1(4.2)$ & $2.93(1.89)$ & & $94.6(3.3)$ & $92.9(3.2)$\\
Swin UNETR $\dagger$\cite{tang2022self} & S &  & $97.9(3.3)$ & $96.6(4.6)$ &  & $92.8(2.7)$ & $92.6(2.3)$ & $92.3(2.4)$ & $92.6(2.7)$ &   & $91.3(3.8)$ & $3.41(2.57)$ & & $94.5(3.4)$ & $93.3(2.7)$\\
Swin UNETRv2 $\dagger$\cite{he2023swinunetr} & S &  & $98.1(3.3)$ & $97.3(4.0)$ &  & $92.7(3.2)$ & $93.2(2.5)$ & $93.4(1.8)$ & $93.1(2.7)$ &   & $91.6(4.3)$ & $2.42(1.19)$ & & $95.5(3.4)$ & $93.1(3.2)$\\
ResUNet34 $\dagger$\cite{he2016deep} & S &  & $98.4(3.5)$ & $97.5(4.5)$ &  & $93.5(2.1)$ & $93.4(2.3)$ & $93.0(2.7)$ & $93.3(2.4)$ &   & $90.6(3.9)$ & $2.19(1.56)$ & & $96.0(2.9)$ & $91.8(3.2)$ \\
VSmTrans $\dagger$\cite{liu2024vsmtrans} & S &  & $98.9(2.3)$ & $97.7(3.6)$ &  & $93.2(2.8)$ & $93.5(1.7)$ & $93.8(1.7)$ & $93.5(2.3)$ &  & $92.1(3.7)$ & $9.06(7.91)$ & & $95.5(3.3)$ & $94.1(2.5)$\\
V-Net $\dagger$\cite{milletari2016v} & S &  & $98.9(2.5)$ & $97.8(3.5)$ &  & $93.7(1.7)$ & $93.8(2.1)$ & $93.2(2.3)$ & $93.5(2.1)$ &  & $92.4(3.6)$ & $1.96(0.70)$ & & $95.9(2.9)$ & $94.0(2.8)$ \\ 
\cmidrule(r){1-2} \cmidrule(lr){4-5} \cmidrule(lr){7-10} \cmidrule(l{0pt}r{2pt}){11-12} \cmidrule(l{2pt}r{3pt}){13-13} \cmidrule(l{3pt}r{2pt}){14-15} \cmidrule(l{2pt}r{0pt}){16-16}
Jang et al.$\dagger$\cite{jang2021fully} \ & I &  & $\underline{96.0(6.2)}$ & - &  & $83.5(1.6)$ & $82.6(2.0)$ & $82.5(1.3)$ & $83.0(1.8)$ &  & $75.6(6.1)$ & $3.07(0.76)$ & & $79.3(3.0)$ & $76.6(5.1)$\\
MWTNet $\dagger$\cite{chen2020automatic} & I &  & $92.6(8.4)$ & - &  & $87.4(1.4)$ & $84.3(2.0)$ & $89.3(1.1)$ & $87.4(2.5)$ &  & $73.9(9.2)$ & $2.29(0.59)$ & & $85.7(4.3)$ & $76.3(6.2)$\\
TSG-GCN $\dagger$\cite{liu2024individual} & S &  & $86.9(9.7)$ & $83.0(11.4)$ &  & $89.3(1.7)$ & $91.0(3.4)$ & $87.8(1.9)$ & $89.2(3.0)$ &  & $76.8(11.2)$ & $2.47(0.76)$ & & $90.1(4.5)$ & $\underline{86.3(4.9)}$\\ 
ToothSeg $\dagger$\cite{cui2021hierarchical} & IS &  & $88.8(10.5)$ & $86.2(7.5)$ &  & $89.3(1.8)$ & $\underline{93.6(0.8)}$ & $89.8(1.7)$ & $90.4(2.6)$ &  & $80.2(11.1)$ & $2.84(1.67)$ & & $91.3(5.1)$ & $81.0(8.0)$\\
SGANet $\dagger$\cite{li2022semantic} & S &  & $92.9(9.6)$ & $\underline{90.8(10.4)}$ &  & $\underline{92.2(1.6)}$ & $92.9(2.5)$ & $\underline{91.9(1.8)}$ & $\underline{92.2(2.1)}$ &  & $\underline{83.8(9.4)}$ & $\underline{2.18(0.74)}$ & & $\underline{94.3(3.2)}$ & $85.7(5.9)$\\ 
\cmidrule(r){1-2} \cmidrule(lr){4-5} \cmidrule(lr){7-10} \cmidrule(l{0pt}r{2pt}){11-12} \cmidrule(l{2pt}r{3pt}){13-13} \cmidrule(l{3pt}r{2pt}){14-15} \cmidrule(l{2pt}r{0pt}){16-16}
\texttt{GEPAR3D} & IS &  & $\mathbf{99.2(2.4)}$ & $\mathbf{98.0(3.7)}$ &  & $\mathbf{95.5(1.2)}$ & $\mathbf{95.1(0.8)}$ & $\mathbf{94.3(1.1)}$ & $\mathbf{95.0(1.4)}$ & & $\mathbf{93.9(3.2)}$ & $\mathbf{1.44(0.70)}$ & & $\mathbf{97.6(1.9)}$ & $\mathbf{95.2(2.1)}$\\
\bottomrule
\end{tabular}%
}
\label{tab:sota_results}
\end{table}%
\noindent
\textbf{Datasets and preprocessing.}
We train and validate our method on a publicly available dataset of 98 CBCT scans~\cite{cui2022fully}, reannotated into 32 classes following the Universal Numbering System~\cite{akram2011micap}.
We test on 46 CBCT scans from 4 medical centers, including two public datasets: Cui et al.~\cite{cui2022fully} and Tooth Fairy 2 (TF2)~\cite{bolelli2024segmenting,cipriano2022improving} (file IDs in accompanying JSON), and 2 in-house sets from a retrospective study (IRB OKW-623/2022) at Polish centers A (11 scans, Carestream CS 9600) and B (9 scans, i-CAT 17-19). To ensure a reliable evaluation of root segmentation, we include test scans where complete roots are fully visible within the field of view. All scans are resampled to an isotropic resolution of $0.4 \times 0.4 \times 0.4$ mm$^{3}$, with Hounsfield Unit intensities clipped to $[0, 5000]$ and normalized to $[0,1]$.\\
\noindent
\textbf{Implementation details.} For training, we randomly crop 128$^3$ patches after extracting the tooth ROI using ground truth labels. The model is trained for 1000 epochs with AdamW, batch size of 2, and a cosine annealing scheduler. The loss function is defined as $L = \Lambda_{1}L_{EDT} + \Lambda_{2}L_{seg} + \Lambda_{3}L_{dir}$, with empirically set weights $\Lambda_{1} = 10$, $\Lambda_{2} = 0.1$, $\Lambda_{3} = 1.0e^{-6}$ for balance. The initial learning rate and weight decay are set to 1e$^{-3}$ and 1e$^{-4}$, respectively. During inference, a lightweight 3D U-Net (patch size: 256$^3$) performs coarse binary segmentation to extract the ROI from the raw CBCT scan. We then apply a sliding window approach (overlap: 0.6) with a Gaussian filter. Our implementation, based on PyTorch 1.13.1 and MONAI 1.3.0, runs on a single NVIDIA A100 GPU.\\
\noindent
\textbf{Evaluation details.} Evaluation includes both multi-class and binary metrics. Multi-class performance is measured with Dice Similarity Coefficient (DSC), Precision (PR), Recall (RC), and Hausdorff Distance (HD). Binary evaluation uses Normalized Surface Dice within a 1-voxel GT boundary ($NSD_1$)~\cite{seidlitz2022robust} and Binary Recall ($RC_B$). While multi-class metrics assess overall performance, binary metrics focus on tooth tissue segmentation completeness, with $RC_B$ highlighting false negatives and $NSD_1$ measuring boundary accuracy, including roots. Instance detection is evaluated via Detection Accuracy (DA) at a 50\% Intersection over Union (IoU) threshold, with instances having \(\text{IoU} > 0.5 \) considered detected. Classification performance is measured using the F1 score. We benchmark \texttt{GEPAR3D} against general segmentation and tooth-specific state-of-the-art methods. General models follow \texttt{GEPAR3D}’s training setup (32-class labels, sliding window inference), while tooth-specific methods are trained per their original protocols, with preprocessing and augmentations matched as closely as possible. Statistical significance is determined via a paired t-test ($p < 0.05$).\\%
\begin{figure}[t!]
\centering
\includegraphics[width=0.99\textwidth]{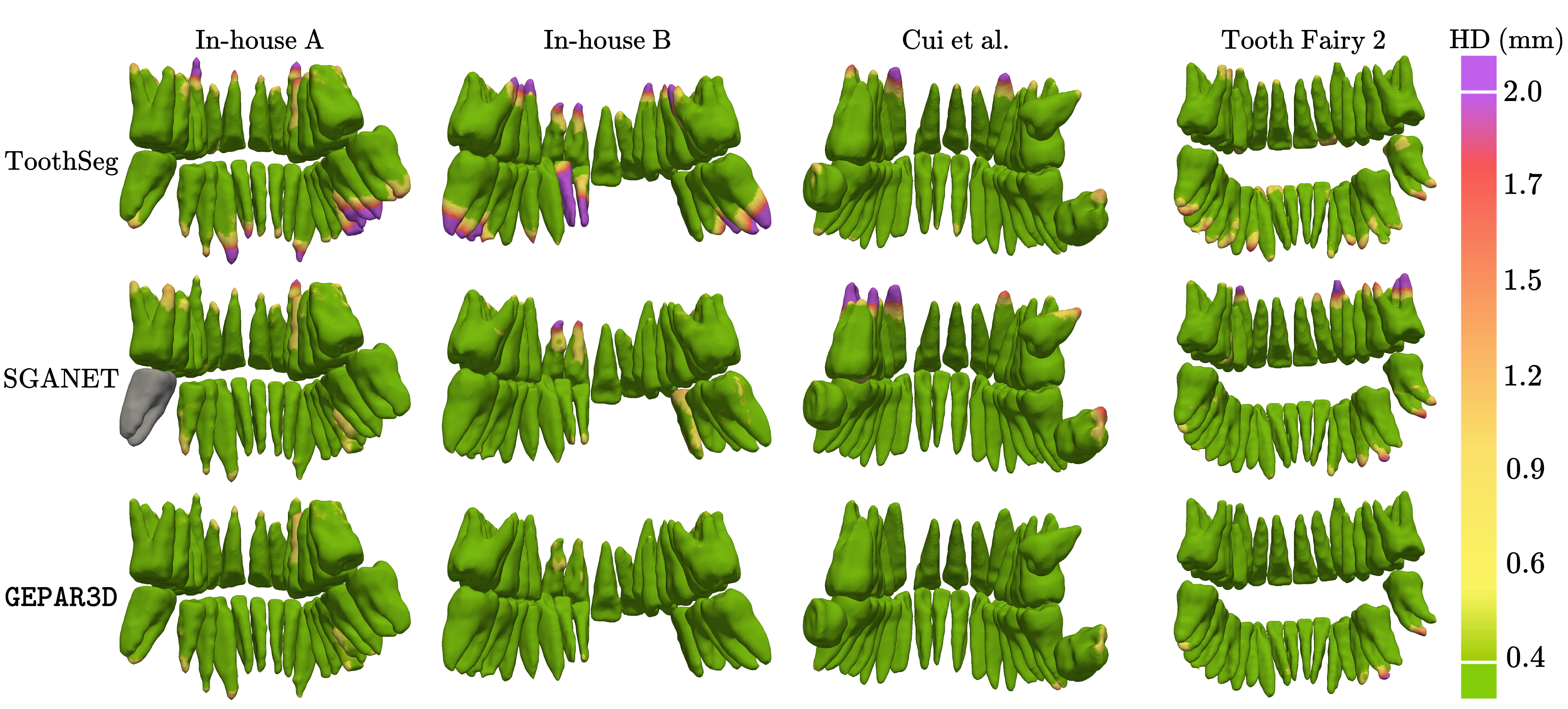}
% \fbox{\rule{0pt}{8cm} \rule{0.95\textwidth}{0pt}}
\caption{Qualitative comparison of \texttt{GEPAR3D} with the two best-performing methods from quantitative results. Surface Hausdorff Distance heatmaps overlaid on GT labels (green = low, purple = high) highlight apex deviations. \texttt{GEPAR3D} shows superior root sensitivity versus tooth-specific baselines. Missing teeth are shown in gray.}
\label{fig:qualitative}
\end{figure}%
\noindent
\textbf{Comparison with state-of-the-art methods.} 
As shown in Table~\ref{tab:sota_results}, \texttt{GEPAR3D} surpasses all competing methods in segmentation and instance detection. It achieves a detection accuracy (DA) of 99.2$\pm$2.4\% (+3.2\% over Jang et al.) and an F1 score of 98.0$\pm$3.7\% (+7.2\% over SGANet), confirming its superior ability to identify and classify tooth instances. This strong detection performance ensures segmentation metrics remain representative, even in challenging cases. \texttt{GEPAR3D} achieves the highest DSC on all external test sets, averaging 95.0$\pm$1.4\% (+2.8\%. over SGANet), alongside the best RC (93.9$\pm$3.2\%) and lowest HD (1.44$\pm$0.50 mm), demonstrating robust generalization. NSD$_1$ of 97.6$\pm$1.9\% (+3.3\% over SGANet) and RC$_B$ of 95.2$\pm$2.1\% (+9.5\% over TSG-GCN), highlight superior tooth tissue completeness. Qualitative results (Fig.~\ref{fig:qualitative}) further validate this, revealing that competing methods often miss substantial root fragments, as shown by per-voxel Hausdorff Distance heatmaps, while \texttt{GEPAR3D} preserves more anatomically complete tooth structures.\\
\noindent
\textbf{Ablation Study.}
We evaluate the impact of various loss functions and network components in \texttt{GEPAR3D} (Table~\ref{tab:architecture_ablation}). As a baseline (\#1), we use Dice+WCE loss. In (\#2), we use the original GWDL with a penalty matrix randomly generated from a uniform distribution to test the loss function's robustness to an uninformative prior (UP). Driven purely by error minimization via optimal transport, shifts the PR–RC balance toward higher sensitivity, resulting in an RC increase of +2.14\% while PR decreases by -3.81\%. Despite the UP, DSC improves (+0.22\%), confirming that loss regularization is not rigid and still enables the model to learn useful representations. Introducing the proposed geometric prior GeoWDL (G) in (\#3) enhances DSC (+1.06\%), improves NSD (+0.09\%), and achieves the highest DA of 99.3\%, demonstrating that structured guidance better aligns with tooth classification, though lowered PR remains. Adding energy map regression (E) in (\#4) via the deep watershed method improves DSC (+1.31\%) and boosts PR (+0.96\%), thereby enhancing focus on tooth instances. In (\#5) we incorporate an auxiliary energy descent direction task (D), yielding further gains in DSC (+0.10\%), RC (+0.29\%), and NSD (+0.16\%), suggesting refined boundary localization. The introduction of GeoWDL in (\#6) increases DSC (+0.28\%) and noticeably improves RC (+1.47\%). Finally, the proposed solution (\#7), which jointly optimizes E and D under G guidance, not only raises PR (+0.36\%) but, more importantly, significantly boosts RC (+3.28\%) over (\#1) and achieves the highest NSD of 97.63\%. Overall, these results demonstrate that proposed components complement each other, enhancing sensitivity in challenging regions.

\begin{table}[t!]\centering
\caption{Ablation study on network and loss components. The best-performing method is highlighted in bold, and the second-best is underlined. DA indicates detection accuracy, PR precision, RC recall and NSD normalized surface dice within 1 voxel boundary. \textbf{G} denotes Geometric Prior loss, \textbf{E} Energy map and \textbf{D} Direction map. $U_{0,1}$ denotes uniform distribution of random cost matrix. $\dagger$ indicates p-value $<$ 0.05.}
\fontsize{9pt}{9pt}\selectfont
\resizebox{\textwidth}{!}{%
\begin{tabular}{@{}lcccccccccc@{}}
\toprule
\cs{\#} & G & E & D & & DSC (\%)$\uparrow$ & PR (\%)$\uparrow$ & RC (\%)$\uparrow$ & $NSD_1$ (\%)$\uparrow$ & DA(\%) $\uparrow$ & F1 (\%)$\uparrow$ \\ 
% \midrule
\cmidrule(r){1-4} \cmidrule(lr){5-9} \cmidrule(l){10-11}
$1^\dagger$ & -          & -       & - & & $93.27(2.40)$ & $94.59(3.73)$ & $90.62(3.85)$ & $95.95(2.94)$ & $98.4(3.5)$ & $97.5(4.5)$ \\
$2^\dagger$ & $(U_{0,1})$ & -       & - & & $93.49(2.35)$ & $90.78(5.31)$ & $92.76(3.53)$ & $95.96(2.83)$ & $99.2(2.0)$ & $98.2(3.3)$  \\ 
$3^\dagger$ & \checkmark  & -       & - & & $94.55(1.49)$ & $91.67(5.39)$ & $92.86(2.35)$ & $96.05(1.94)$ & $\mathbf{99.3(1.9)}$ & $\mathbf{98.4(3.2)}$ \\ 
\cmidrule(r){1-4} \cmidrule(lr){5-9} \cmidrule(l){10-11}
$4^\dagger$ & - & \checkmark          & - & & $94.58(1.28)$ & $\mathbf{95.55(2.94)}$ & $91.36(3.67)$ & $97.25(1.93)$ & $99.2(2.2)$ & $98.0(3.5)$ \\
$5^\dagger$ & - & \checkmark & \checkmark  &  & $94.68(1.13)$ & $95.52(2.86)$ & $91.65(3.81)$ & $97.41(1.83)$ & $99.2(2.4)$ & $98.0(3.5)$\\ 
$6^\dagger$ & \checkmark & \checkmark & -  &  & $94.96(1.13)$ & $94.33(3.24)$ & $93.12(3.33)$ & $97.58(1.90)$ & $99.1(2.5)$ & $97.9(3.6)$\\
7 & \checkmark & \checkmark & \checkmark  &  & $\mathbf{95.01(1.36)}$ & $94.95(3.45)$ & $\mathbf{93.90(3.18)}$ & $\mathbf{97.63(1.94)}$ & $99.2(2.4)$ & $98.0(3.7)$ \\
\bottomrule
\end{tabular}%
}
\label{tab:architecture_ablation}
\end{table}
\section{Conclusions}
We present \texttt{GEPAR3D}, which combines geometric prior-assisted learning with deep watershed instance detection to improve tooth segmentation, particularly for fine root structures. Extensive experiments demonstrate its superiority over state-of-the-art methods, with enhanced segmentation supporting better orthodontic planning and root resorption assessment. We ensure reproducibility by validating on public datasets, sharing code and implementation details. However, our study has limitations. Training was restricted to adult teeth, which may limit applicability to younger patients. Additionally, while geometric prior loss is crucial for encoding anatomical constraints, it can be overly sensitive when used alone, requiring careful tuning. In \texttt{GEPAR3D}, its integration with instance regression balances sensitivity and precision, mitigating this issue. While our method significantly improves root segmentation, further gains could be achieved with larger datasets and self-supervised training. Finally, resorption analysis requires comparing sequential scans to a reliable baseline segmentation. As no public CBCT datasets include resorbed annotations, we focused on validating apex segmentation accuracy, since under-segmentation would mask subsequent root shortening. To conclude, this work underscores the importance of root segmentation and aims to inspire future research.

\begin{credits}
\subsubsection{\ackname} This work is supported by the EU's Horizon 2020 programme (grant no. 857533, Sano) and the Foundation for Polish Science's International Research Agendas programme (MAB PLUS/2019/13), co-financed by the EU under the European Regional Development Fund and the Polish Ministry of Science and Higher Education (contract no. MEiN/2023/DIR/3796). This research was funded in whole or in part by National Science Centre, Poland 2023/49/N/ST6/01841. For the purpose of Open Access, the author has applied a CC-BY public copyright licence to any Author Accepted Manuscript (AAM) version arising from this submission.
\subsubsection{\discintname}
The authors have no competing interests to declare.
\end{credits}

\bibliographystyle{splncs04}
\bibliography{mybibliography}

% \end{document}

\appendix
\counterwithin{figure}{section}
\counterwithin{table}{section}

%_______________________________________________________________
\section{GEPAR3D Dataset}

We train and validate our proposed method using a publicly available dataset comprising 98 Cone-Beam Computed Tomography (CBCT) scans~\cite{cui2022fully}. To standardize the annotation schema, we redefined the original instance labels into 32 distinct classes based on the widely accepted Universal Dental Notation system. For external testing and generalization assessment, we evaluate the performance of GEPAR3D on an independent set of 46 CBCT scans collected from four different medical centers. This external test set includes data from two publicly available datasets~\cite{cui2022fully,bolelli2024segmenting,cipriano2022improving} as well as two proprietary in-house datasets. This diverse test set allows us to robustly assess the cross-domain adaptability and real-world applicability of our approach.

\begin{figure}[h]
\centering
\includegraphics[width=0.6\textwidth]{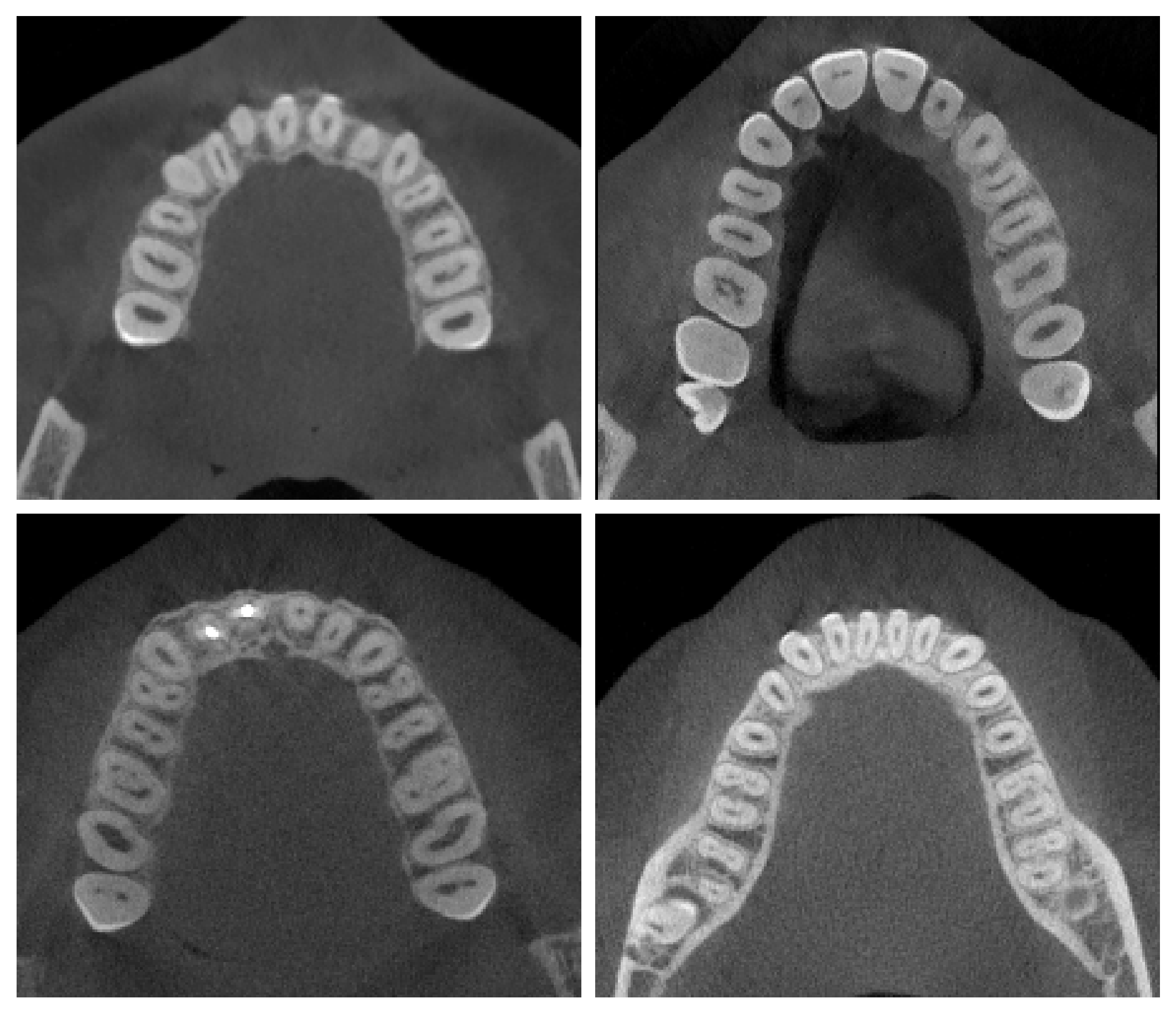}
\caption{Samples from our in-house CBCT scans test set are shown: the first row represents scans from center A, and the second from center B. We release these scans for the community to enhance research in the field.}
\label{fig:GEPAR3D_dataset_slices}
\end{figure}

%_______________________________________________________________
\subsection{Re-annotated Training Dataset}

\begin{figure}[h]
\centering
\includegraphics[width=0.6\textwidth]{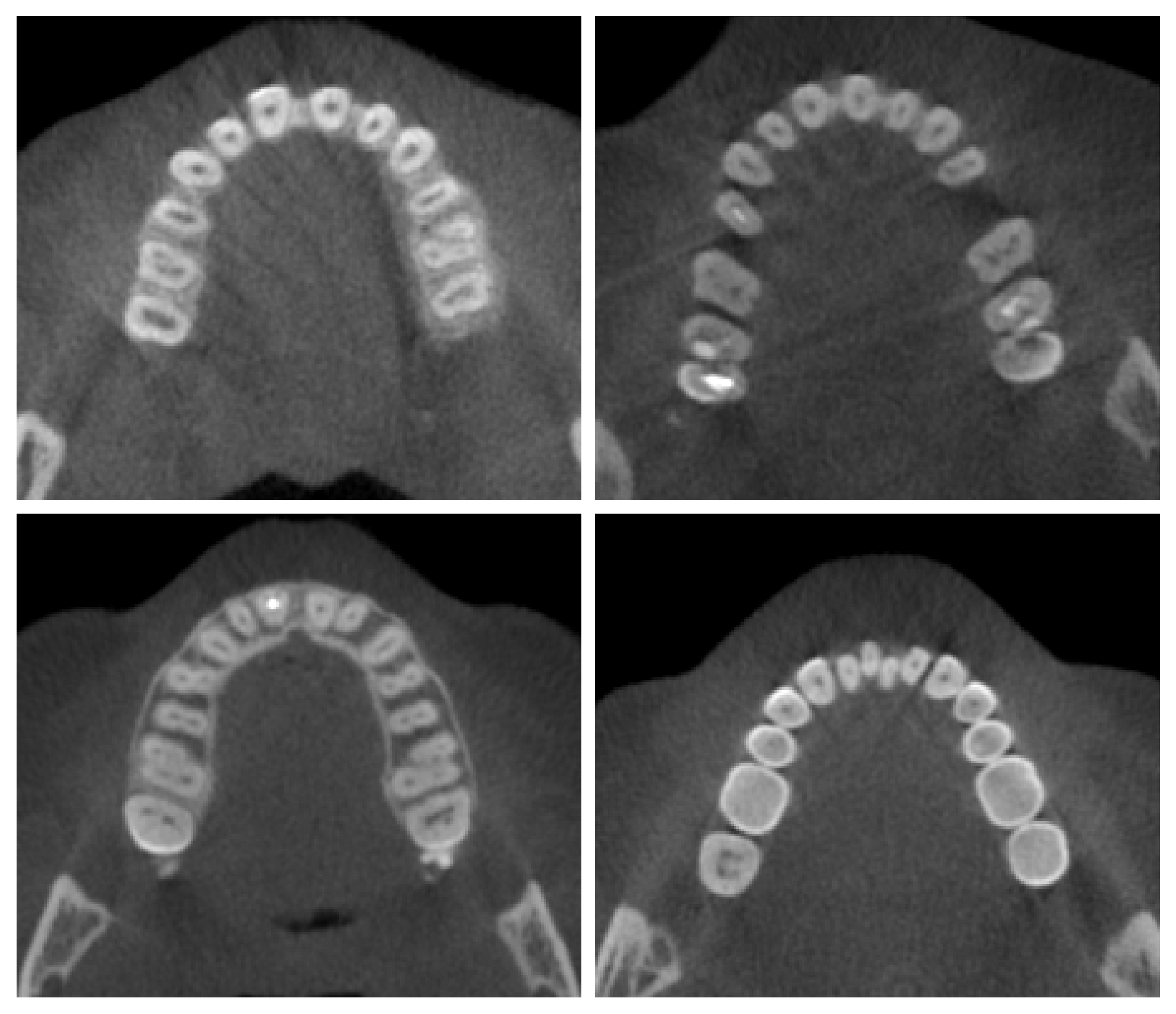}
\caption{Samples of publicly available CBCT dataset we use for training. We annotate its tooth instances into 32 classes to supervise the model with morphologically meaningful labels, which we make publicly available.}
\label{fig:training_cbct_slices}
\end{figure}

\begin{figure}[!h]
\centering
\includegraphics[width=\textwidth]{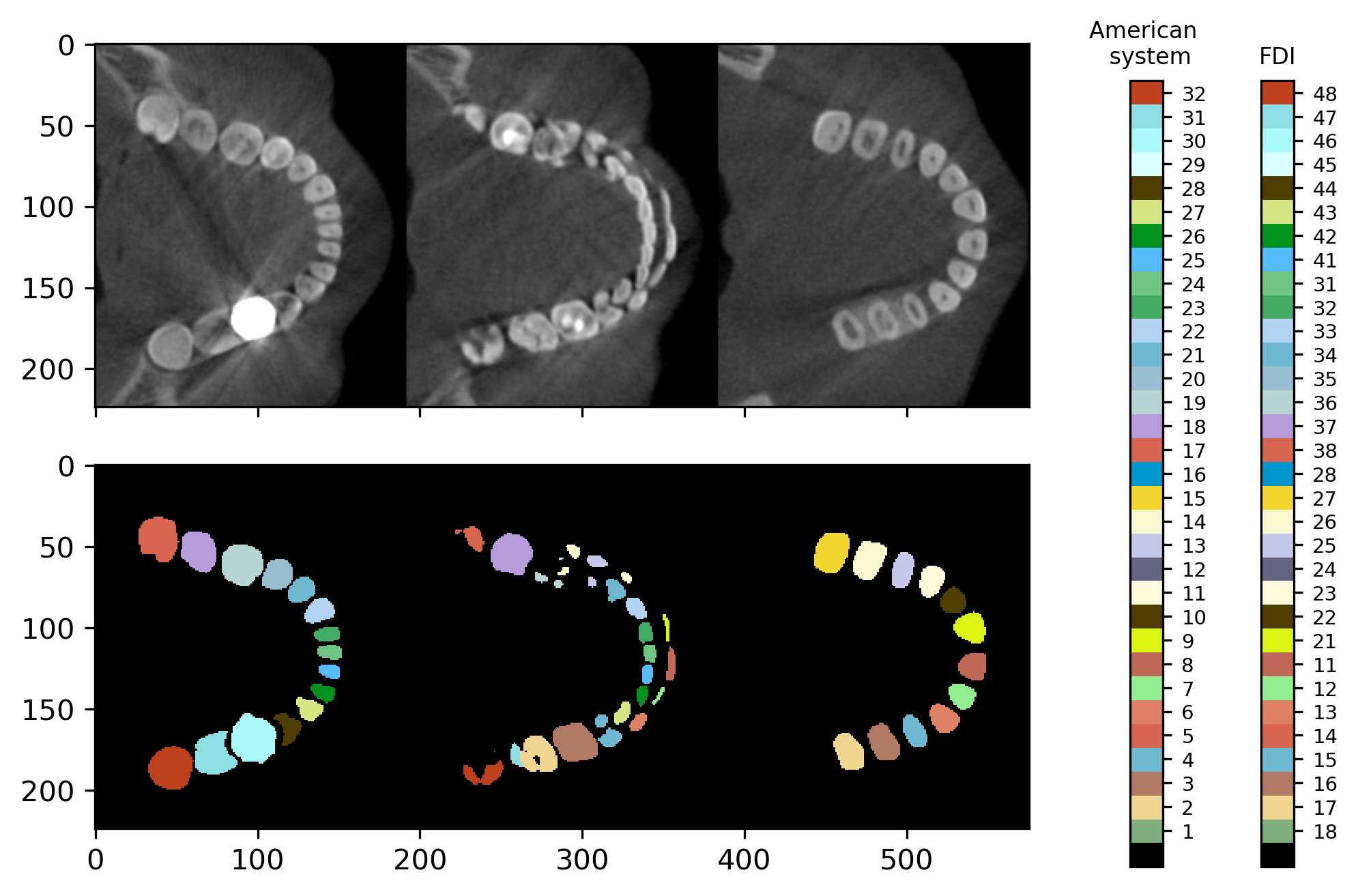}
\caption{Training dataset sample. We annotated the data into 32 classes according to FDI and American Systems. Sections XY, from left to right, display the mandible, occlusion (bite), and maxilla.}
\label{fig:china_systems}
\end{figure}

\begin{figure}[!h]
\centering
\includegraphics[width=\textwidth]{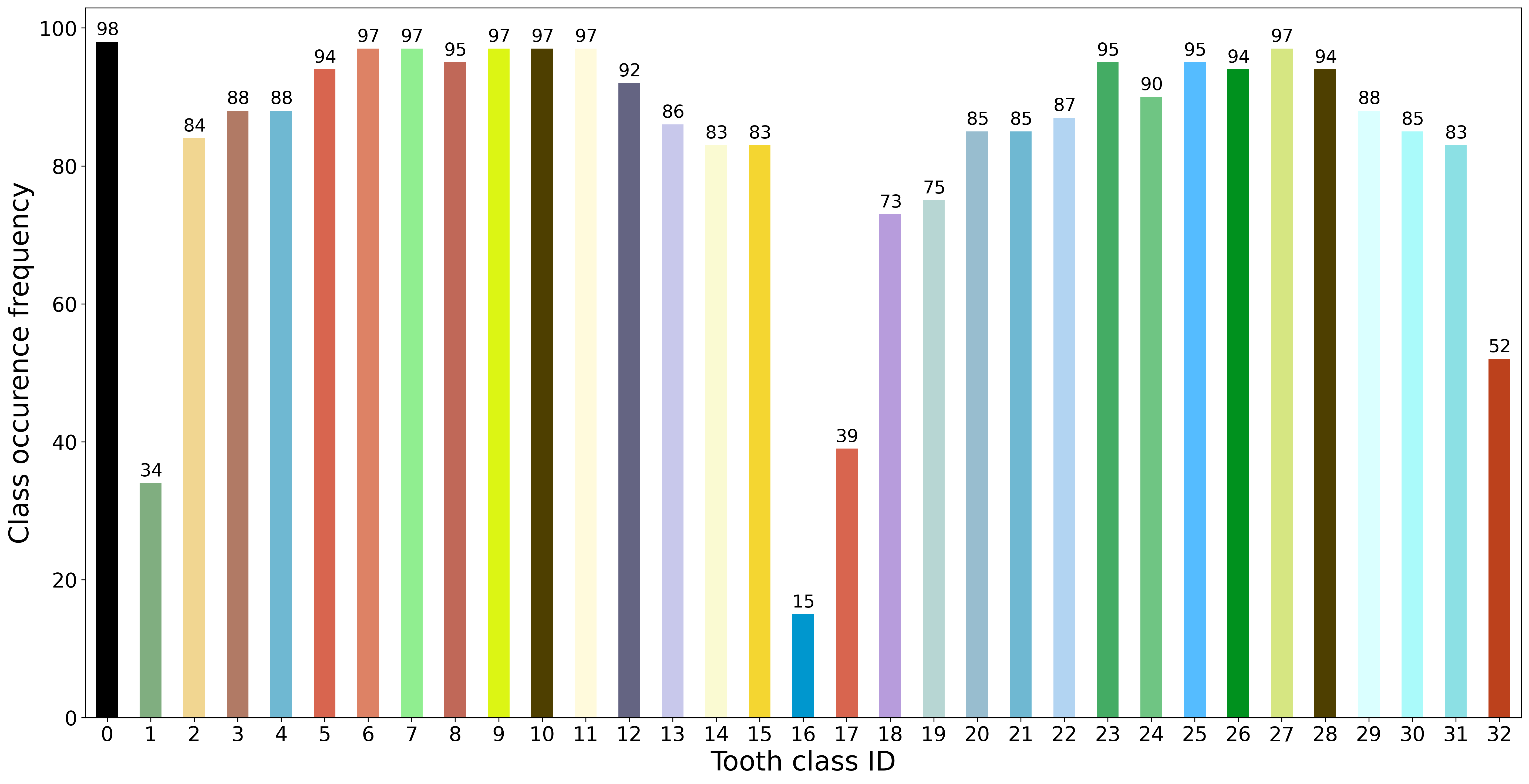}
\caption{Training dataset tooth class frequencies reveal a significant under-representation of third molars, which poses a notable challenge for accurate segmentation. This difficulty is compounded by the high morphological variability of third molars, including frequent deviations in position, angulation, and rotation. Additionally, third molars are often partially or fully retained within the alveolar bone, further complicating their identification and delineation in CBCT scans.}
\label{fig:training_occurence}
\end{figure}

We have annotated a publicly available dental CBCT dataset into 32 distinct classes (as shown in Fig. \ref{fig:china_systems}) to provide morphologically meaningful labels that can be used for model supervision. These annotated labels are made publicly available for the research community, and can be accessed at \href{https://zenodo.org/records/15739014}{\texttt{zenodo.org/records/GEPAR3D}}. Researchers interested in obtaining the corresponding raw CBCT scans should contact the original data providers, as detailed in \cite{cui2022fully}. Fig. \ref{fig:training_cbct_slices} presents a selection of scan samples from the dataset, while Fig. \ref{fig:training_occurence} illustrates the frequency of class occurrences across the dataset. Due to the marked class imbalance, with third molars being significantly under-represented, diminished segmentation accuracy for this class is expected. Some scans contain deciduous teeth, but in our approach, focusing on normal adult dentition, they are excluded at the preprocessing stage.

\subsection{In-house Test Set}

We utilize 20 in-house CBCT scans collected from two medical centers, referred to as Center~A and Center~B (scan examples are shown in Fig.~\ref{fig:GEPAR3D_dataset_slices}). Both datasets come from Warsaw, Poland. The data consist of anonymized CBCT volumes and voxel-wise segmentation masks for 3D tooth structures. The use of the data was approved by the Institutional Review Board (IRB Approval ID: OKW-623/2022). Center~A provides 11 scans, while Center~B contributes 9 scans. The scans were acquired using the \textit{Carestream CS 9600} and \textit{i-CAT 17-19} imaging systems, with slice thicknesses of $0.15$~mm/px and $0.2$~mm/px, respectively.

The ground truth annotations for the test set were performed by a board-certified orthodontist with 5 years of clinical experience and independently verified by a senior orthodontist with 25 years of clinical practice, ensuring a high level of annotation reliability. All scans were resampled to an isotropic resolution of $0.4 \times 0.4 \times 0.4~\text{mm}^3$, aligning with the resolution of the publicly available training dataset. The field of view (FoV) of the scans is restricted to the region of interest (ROI) containing the dentition.

\begin{figure}[!h]
\centering
\includegraphics[width=\textwidth]{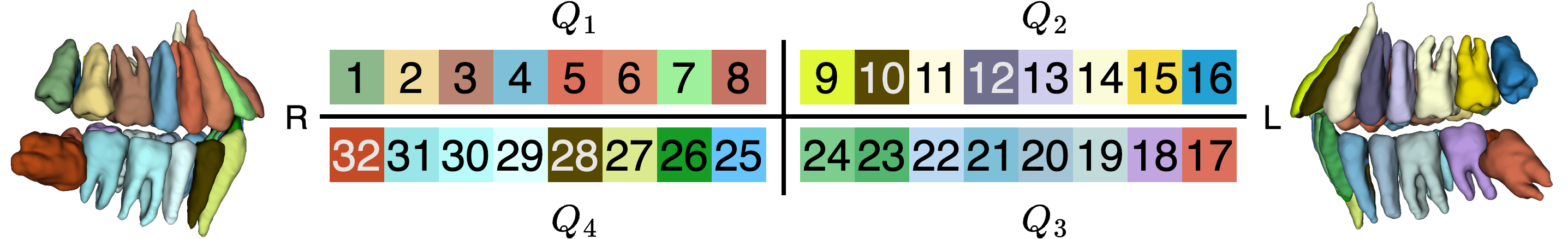}
\caption{The Universal Numbering System, also called the American System. Tooth number 1 is the maxillary right third molar, with the count progressing along the upper arch to the left side. The numbering then resumes at the mandibular left third molar, number 17, continuing along the lower teeth to the right side.}
\label{fig:american_system}
\end{figure}

The dataset is annotated according to the dental notation system into 32 classes, and we present it in the Universal Numbering System, as shown in Fig. \ref{fig:american_system}. We associate the colormap with class IDs to assist the reader in interpreting the classification results.

\FloatBarrier
%_______________________________________________________________
\section{Teeth ROI Detection}

\begin{figure}[t]
\centering
\includegraphics[width=\textwidth]{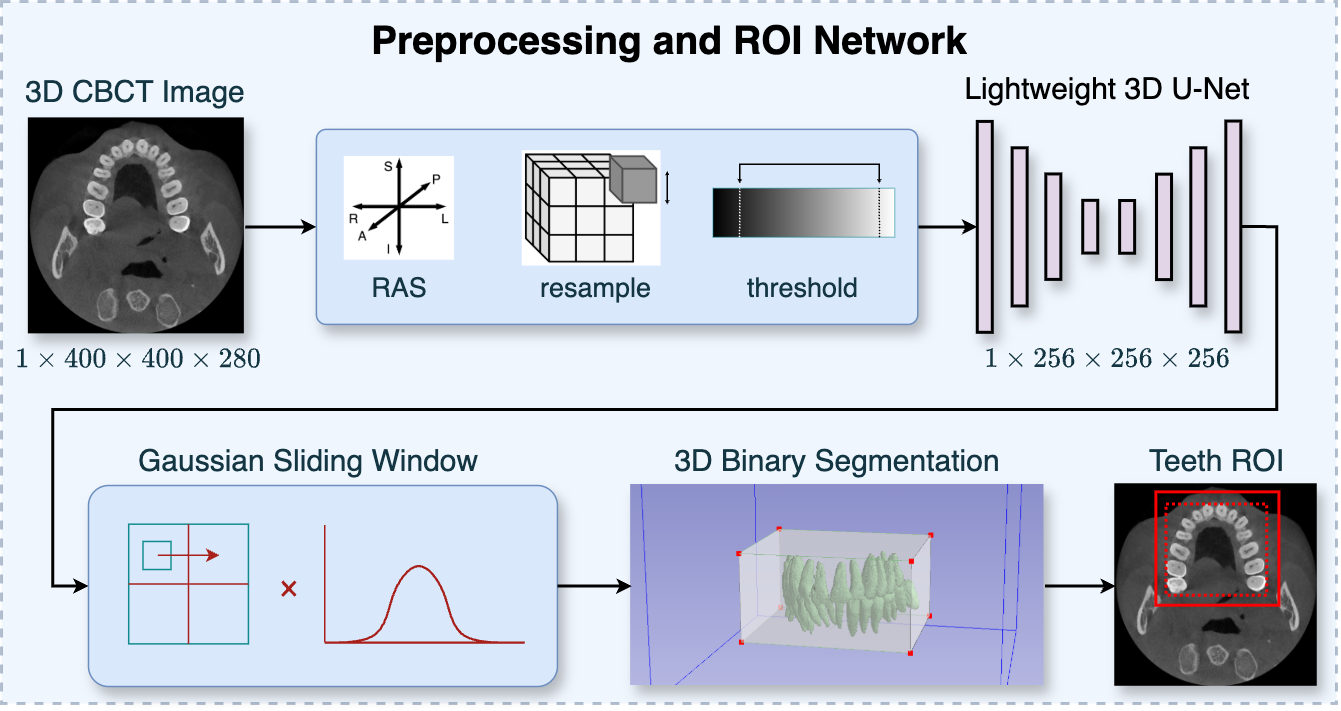}
\caption{An overview of the initial preprocessing, the preliminary stage of our method, is provided. Each CBCT scan input is converted to the RAS (Right, Anterior, Superior) orientation system, resampled to a $0.4 \times 0.4 \times 0.4 mm^{3}$ isotropic resolution, and has its Hounsfield Units (HU) clipped to the range [$0, 5000]$, and finally normalized between $[0, 1]$. The input scan is then fed into the lightweight UNet 3D network for coarse segmentation of the teeth's Region of Interest (ROI). For training time, random patches of size $256 \times 256 \times 256$ are selected from the raw CBCT scan. During inference, the Gaussian Sliding Window algorithm is used to perform 3D binary segmentation of the whole scan. Based on the result, an ROI around the teeth with a small additional margin is cropped.}
\label{fig:preprocessing_steps}
\end{figure}

To prepare CBCT scans (\textit{C$\times$H$\times$W$\times$D}) for automated analysis, a series of carefully designed preprocessing steps are applied to enhance consistency and model performance (see Fig. \ref{fig:preprocessing_steps}). First, volumes are aligned using the RAS (Right, Anterior, Superior) coordinate system to ensure anatomical correspondence across different patients. Spatial resampling to an isotropic resolution of $0.4 \times 0.4 \times 0.4~\text{mm}^3$ follows, which preserves geometrical fidelity and simplifies downstream 3D processing. Intensity normalization is then performed by clipping Hounsfield Unit (HU) values into the range $[0, 5000]$, mitigating the influence of artifacts, and scaling them to the $[0, 1]$ range, making the data more suitable for neural network training. While the Hounsfield Unit (HU) values for dental tissues, such as enamel, typically reach a maximum of around $3000$ HU, extending the intensity range improves the network's robustness in the presence of metallic artifacts. This standardized input is passed into a compact 3D U-Net, whose role is to identify a coarse spatial envelope around the teeth, enabling efficient localization. The coarse U-Net is a 4-stage deep encoder-decoder architecture for binary segmentation, using strided convolutions and transposed convolutions at the start of each block for downsampling and upsampling, respectively. It employs ReLU activation, Instance Normalization, and additive skip connections for efficiency. Transposed convolutions, used in place of fixed interpolation, ensure that the output resolution matches the input while allowing a learnable, data-driven upsampling. During training, randomly sampled patches of size $256 \times 256 \times 256$ are used to improve generalization while managing the computational load. At inference time, the full scan is segmented using a Gaussian Sliding Window strategy, which smoothly integrates predictions across overlapping regions. From this coarse output, a bounding volume is extracted, slightly expanded to ensure full coverage of dental structures (see Fig. \ref{fig:preprocessing_steps} 3D Binary Segmentation), which serves as the focused Region of Interest (ROI) for the next stage of the pipeline.  

\FloatBarrier
%_______________________________________________________________
\section{Statistical Shape Model}
% \lipsum[1-2]

\begin{figure}[!h]
\centering
\includegraphics[width=0.6\textwidth]{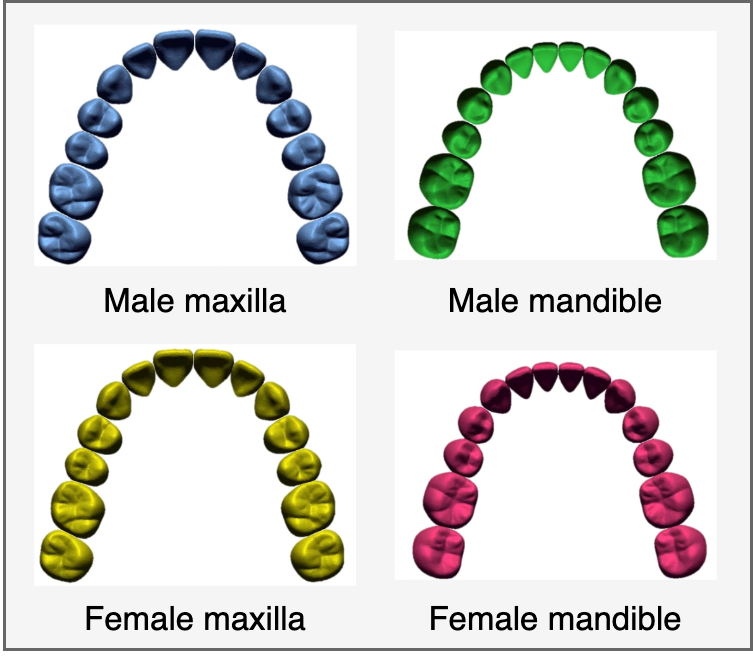}
\caption{Male and female statistical shape model (SSM) of normal dentition created by (Kim et al. 2022).}
\label{fig:SSM_male_female}
\end{figure}

To incorporate anatomical priors into root segmentation, we integrate a high-resolution Statistical Shape Model (SSM) of normal dentition developed by Kim et al. (2022) \cite{kim2022developing}. This SSM encodes detailed morphological and spatial variability of teeth in a compact, data-driven representation derived from a population of individuals with a clinically defined normal dentition.

The SSM was constructed from dental casts collected between 1997 and 2019 using alginate impressions~\cite{imbery2010accuracy} to ensure methodological consistency. Inclusion criteria were strictly defined to isolate normal dentition morphology: complete permanent dentition (excluding third molars), absence of prosthetics or orthodontic interventions, minimal crowding/spacing, and no morphological abnormalities. The cohort was restricted to Korean individuals aged 15 to 30 years to control for ethnic and age-related variation in arch form. After expert review by two experienced orthodontists, the final model was based on 47 male subjects (mean age $20.3 \pm 4.1$) and 37 female subjects (mean age $19.3\pm3.7$). The SSMs authors constructed separate models for male and female dentition to account for known sexual dimorphism in arch width and tooth dimensions (Fig.~\ref{fig:SSM_male_female}). However, since our training dataset lacks patient sex information, we use an averaged representation of male and female models to generate population-based centroid positions.   

\begin{figure}[!h]
\centering
\includegraphics[width=0.5\textwidth]{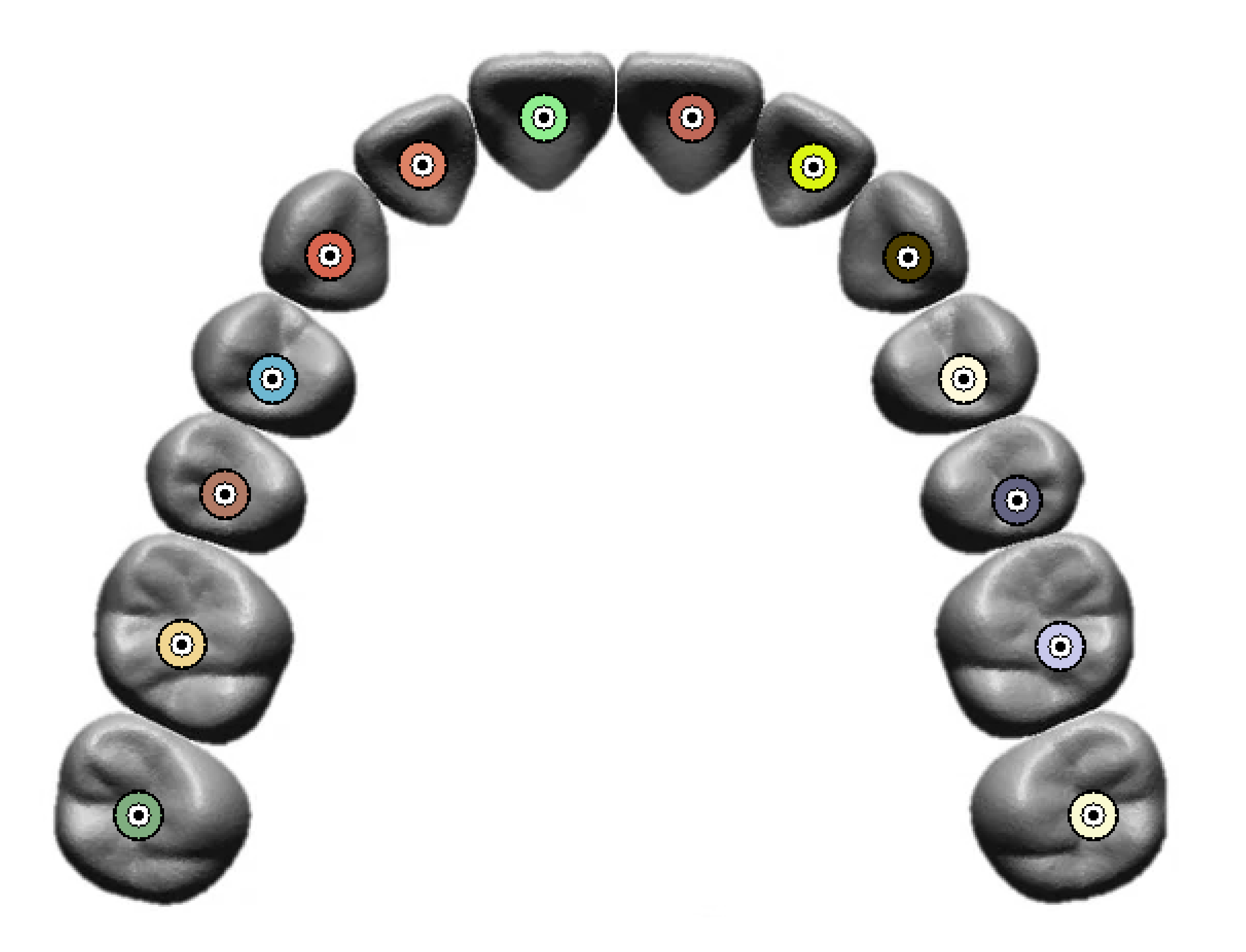}
\caption{We use high-resolution SSM to obtain statistical tooth positions. We separate the model into tooth instances and for each, we determine the centroid.}
\label{fig:SSM_centroids}
\end{figure}

\begin{figure}[!h]
\centering
\includegraphics[width=0.9\textwidth]{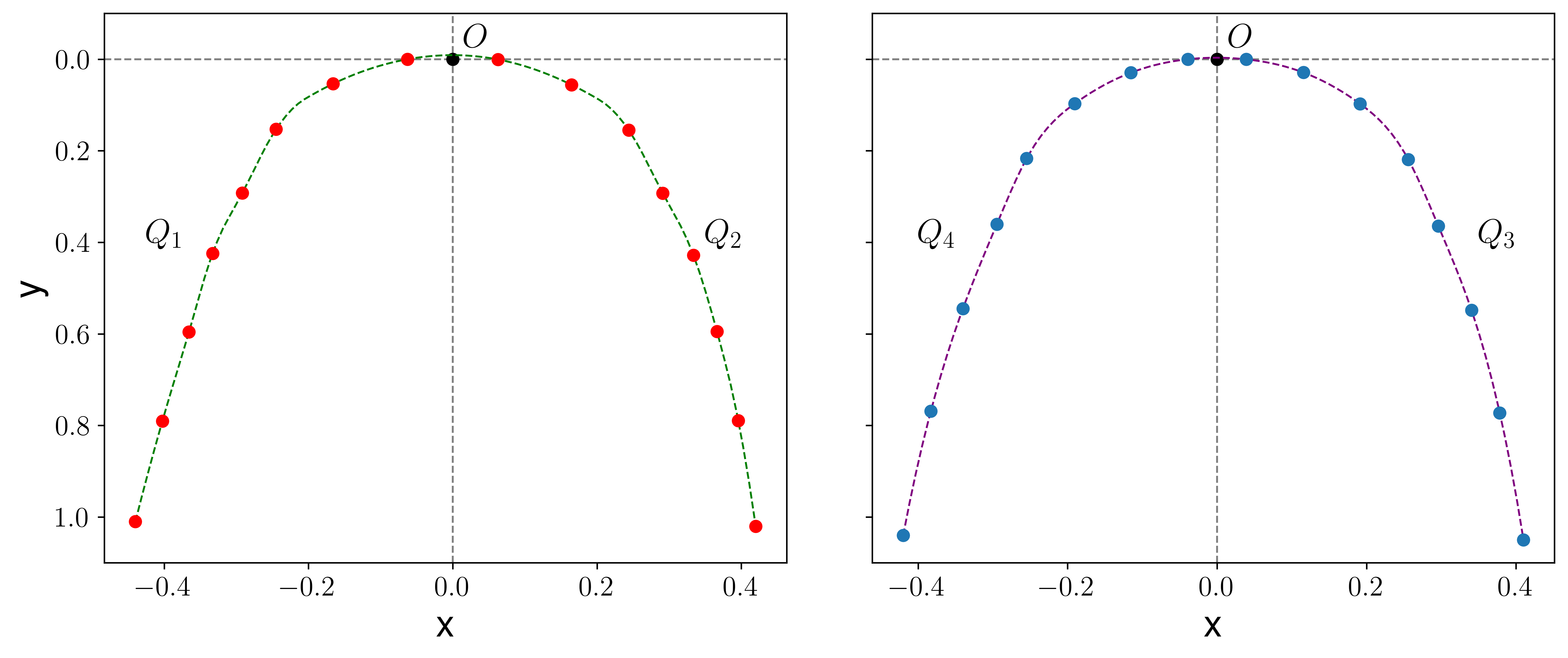}
\caption{SSM-based centroids in normalized coordinate system with interpolated third molar. $Q_k$ denotes oral cavity quadrant and $O$ origin.}
\label{fig:SSM_cooridantes}
\end{figure}

We utilize the SSM to extract statistically representative tooth positions, which serve as geometric priors in our segmentation pipeline. The model is decomposed into individual tooth instances, and the centroid of each tooth is computed to define its canonical location in a normalized coordinate space (Fig.\ref{fig:SSM_centroids}). The third molars, which were excluded from the original model, are interpolated based on adjacent molar geometry to maintain anatomical completeness (Fig.\ref{fig:SSM_cooridantes}).

The integration of the SSM improves segmentation robustness by incorporating high-level anatomical consistency not captured by intensity cues alone. Unlike heuristic constraints or hand-crafted rules, the SSM encodes population-derived 3D morphology, reducing sensitivity to local noise or occlusion artifacts. Its objectivity and reproducibility further ensure reliable behavior across variable imaging conditions, especially when applied to external data, mimicking the practical application of the trained model.

\FloatBarrier
%_______________________________________________________________
\section{SSM-based Wasserstein Matrix}
% \lipsum[1]

\begin{figure}[!h]
\centering
\includegraphics[width=\textwidth]{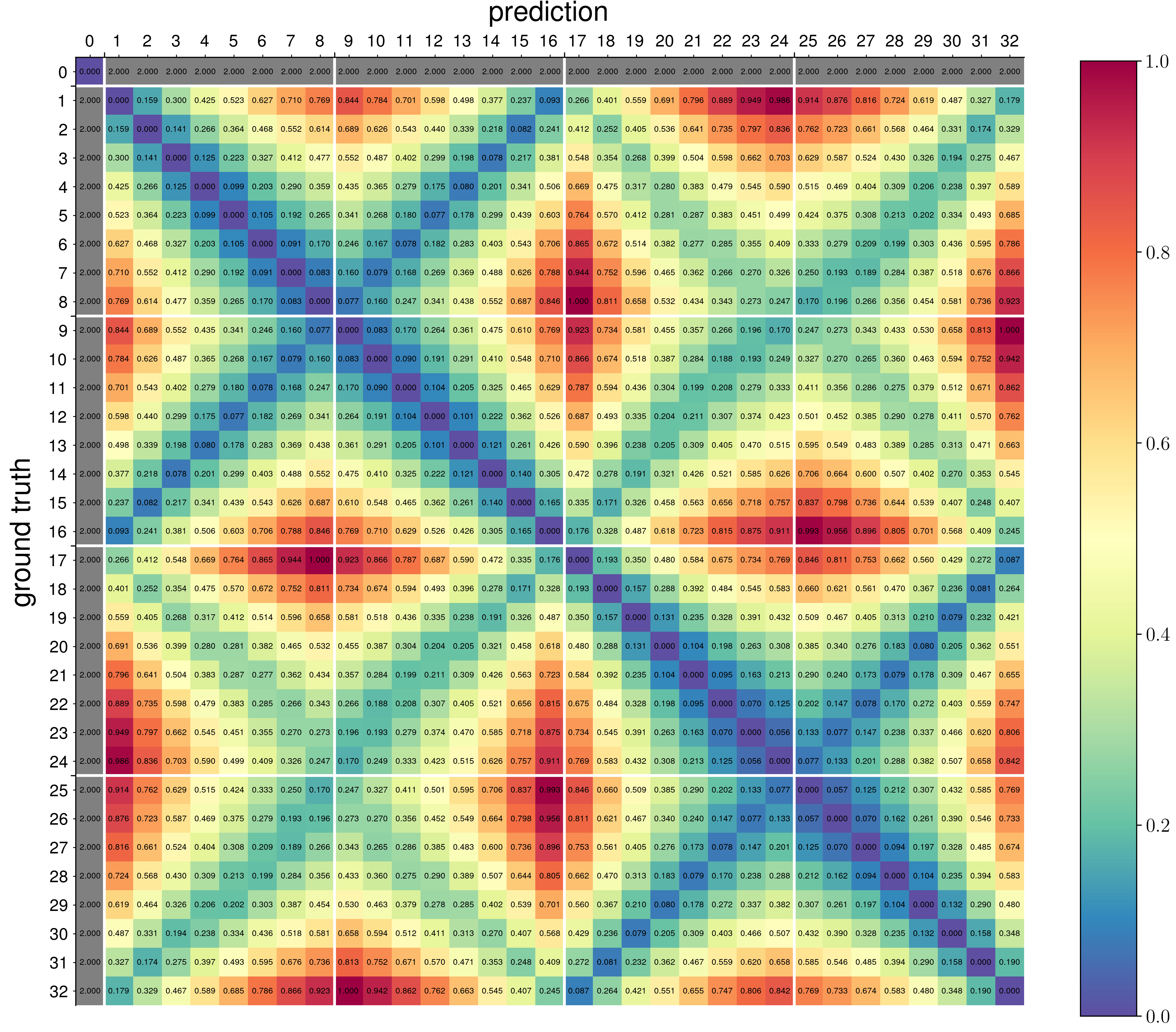}
\caption{Geometrical Wasserstein Distance (GWD) loss - Wasserstein matrix based on geometrical and morphological priors. The colormap displays penalty values normalized to 1, with red indicating the highest penalty. In addition to the main diagonal, perpendicular valleys of low penalties can be observed. They correspond to directly adjacent teeth in the opposite arch or are located symmetrically within the same dental arch. The former are larger due to greater morphological differences.}
\label{fig:wasserstein_matrix_2d}
\end{figure}

To construct the Geometrical Wasserstein Distance (GWD) loss, we derive a Wasserstein matrix encoding the geometric and morphological relationships between tooth classes (32 classes and background) based on the Statistical Shape Model (SSM). The visualization of the 2D matrix (Fig.~\ref{fig:wasserstein_matrix_2d}) shows the penalty values normalized to one, with red indicating the highest misclassification cost. Notably, misclassifying tooth voxels as background incurs the most severe penalty, set to 2, to strongly discourage false negatives, which is particularly critical for accurately segmenting root apex voxels. Furthermore, low-penalty values appear along the main diagonal, corresponding to correct class assignments, and in perpendicular bands reflecting adjacent teeth either in the opposite arch or symmetrically located within the same arch. For example, penalties between neighboring molars or between contralateral incisors are lower due to morphological and spatial similarity, whereas distant or dissimilar teeth, such as a canine vs. a third molar, incur high penalties. The interplay between geometric proximity and morphological similarity is reflected in the penalty structure of the Wasserstein matrix. For example, misclassifying tooth 1 as tooth 16, both third molars, incurs a relatively low penalty of $0.093$ due to high morphological similarity, despite their spatial separation. In contrast, confusing tooth 1 with tooth 2 (a second molar), which is anatomically adjacent, results in a higher penalty of $0.159$, as second molars differ more in shape. Interestingly, misclassifying tooth 1 as tooth 32 (a mandibular third molar) yields an even higher penalty of $0.179$, despite tooth 32 being closer in space than tooth 2. This is due to the morphological dissimilarity between maxillary and mandibular teeth, which the loss function captures by assigning stronger penalties across dental arches.

\begin{figure}[!h]
\centering
\includegraphics[width=0.6\textwidth]{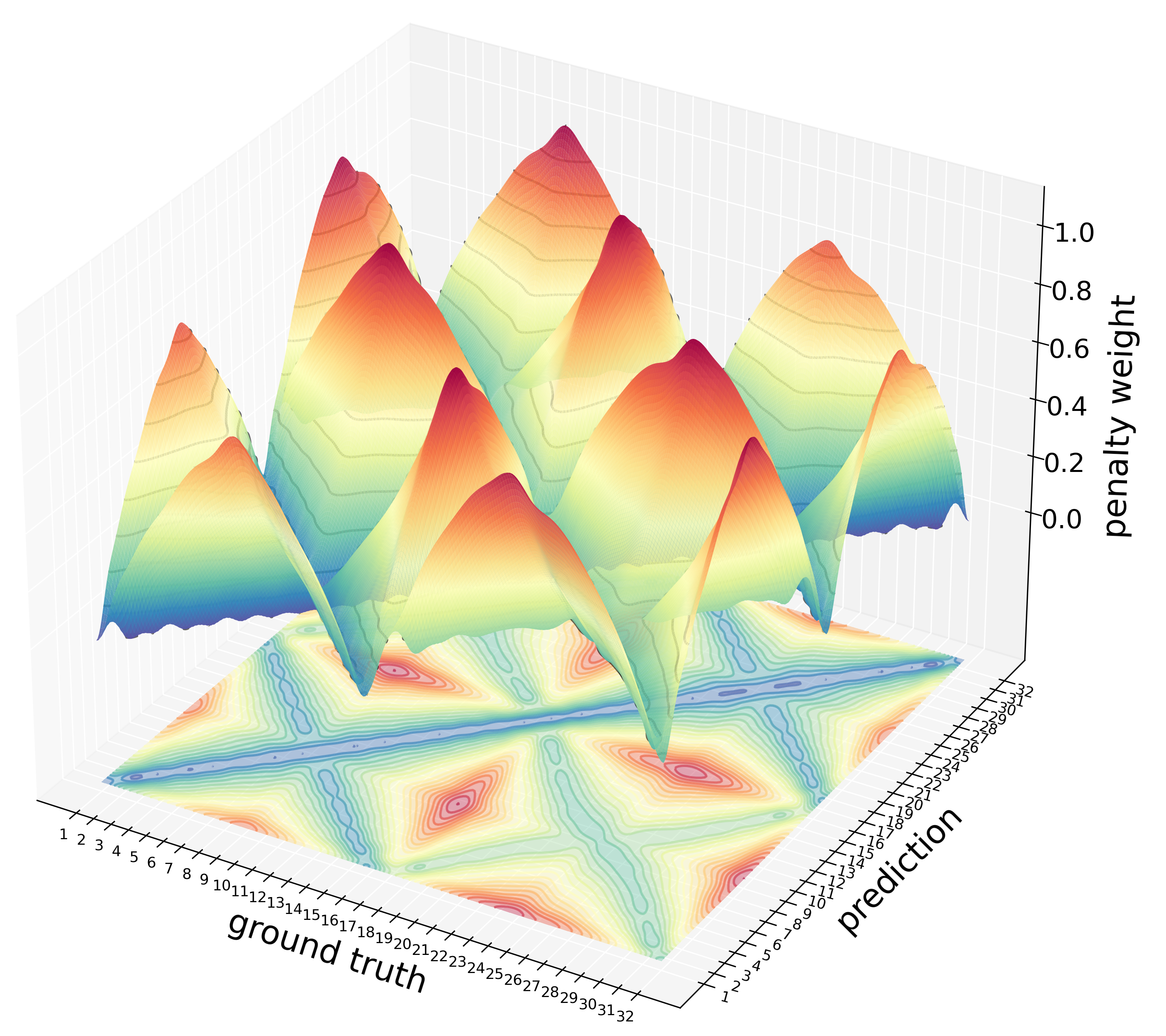}
\caption{Wasserstein matrix based on geometrical and morphological priors used within Geometrical Wasserstein Distance (GWD) loss.}
\label{fig:wasserstein_matrix_3d}
\end{figure}

Fig.~\ref{fig:wasserstein_matrix_3d} presents the Wasserstein matrix embedded in 3D space, illustrating how spatial geometry and tooth morphology jointly inform class proximity. The surface formed by penalty values reveals a structured landscape, with valleys corresponding to morphologically and spatially similar teeth. This can be interpreted analogously to an optimization landscape, where regions of low penalty represent anatomically plausible class assignments, local minima in the loss space, toward which the model is implicitly guided during training. The GWD loss thus not only penalizes implausible misclassifications but also shapes the learning dynamics by favoring anatomically coherent predictions, helping the model approximate a biologically consistent minimum in the tooth classification space.

\begin{figure}[!t]
\centering
\includegraphics[width=\textwidth]{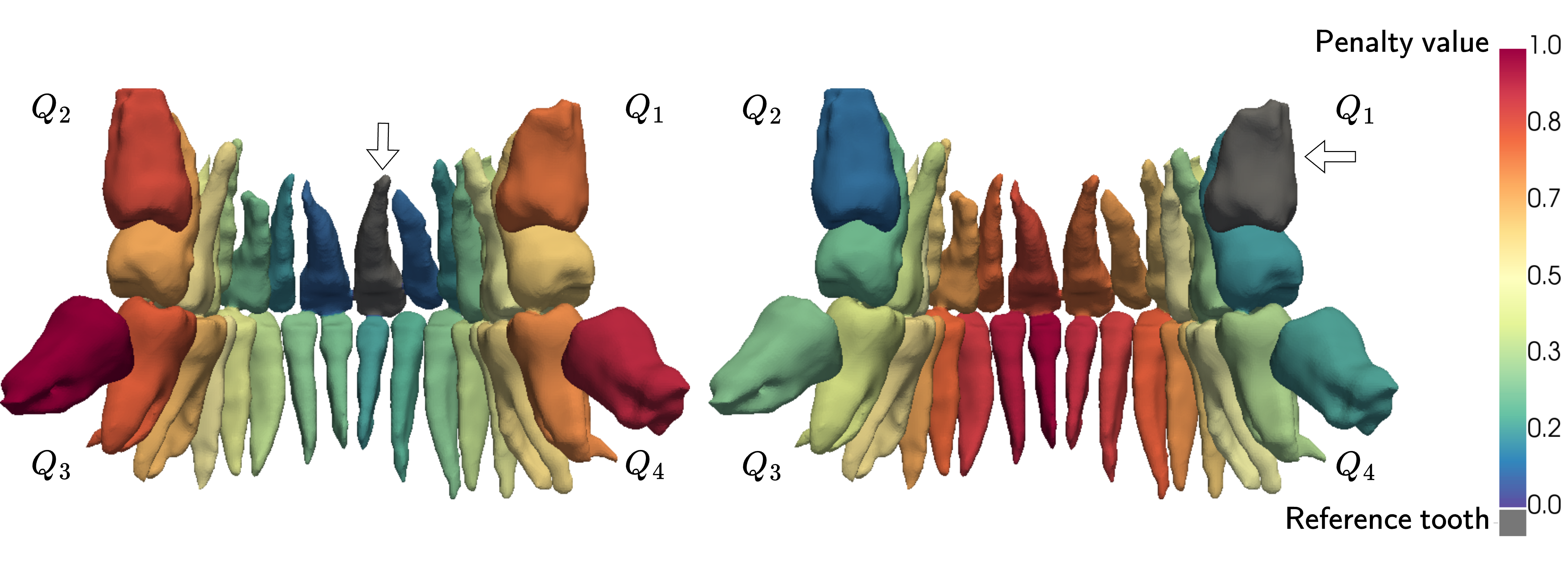}
\caption{Intermediate step in the computation of the Wasserstein loss, where penalties are assigned to each class based on the Wasserstein distance from a designated reference tooth. A 3D heatmap visualization, overlaid on the segmentation GT label, is shown for two different reference teeth (indicated by arrows and gray overlays). The highest misclassification penalties correspond to teeth that are both morphologically dissimilar and spatially distant from the reference.}
\label{fig:wasserstein_3d_reference}
\end{figure}

Figure~\ref{fig:wasserstein_3d_reference} further demonstrates the application of this matrix during training. Here, we visualize a 3D heatmap of misclassification penalties overlaid on ground truth segmentations for two reference teeth. Misclassifications involving teeth adjacent or symmetrical to the reference exhibit moderate penalties, while those involving distant or morphologically dissimilar teeth are heavily penalized. While only two reference teeth are shown for illustration, in practice, such a matrix is instantiated for each tooth class during loss computation to assess the cost of probability transport across the full distribution of predicted labels.

\FloatBarrier
%_______________________________________________________________
\section{Deep Watershed}
% \lipsum[1]

\begin{figure}[!h]
\centering
\includegraphics[width=0.6\textwidth]{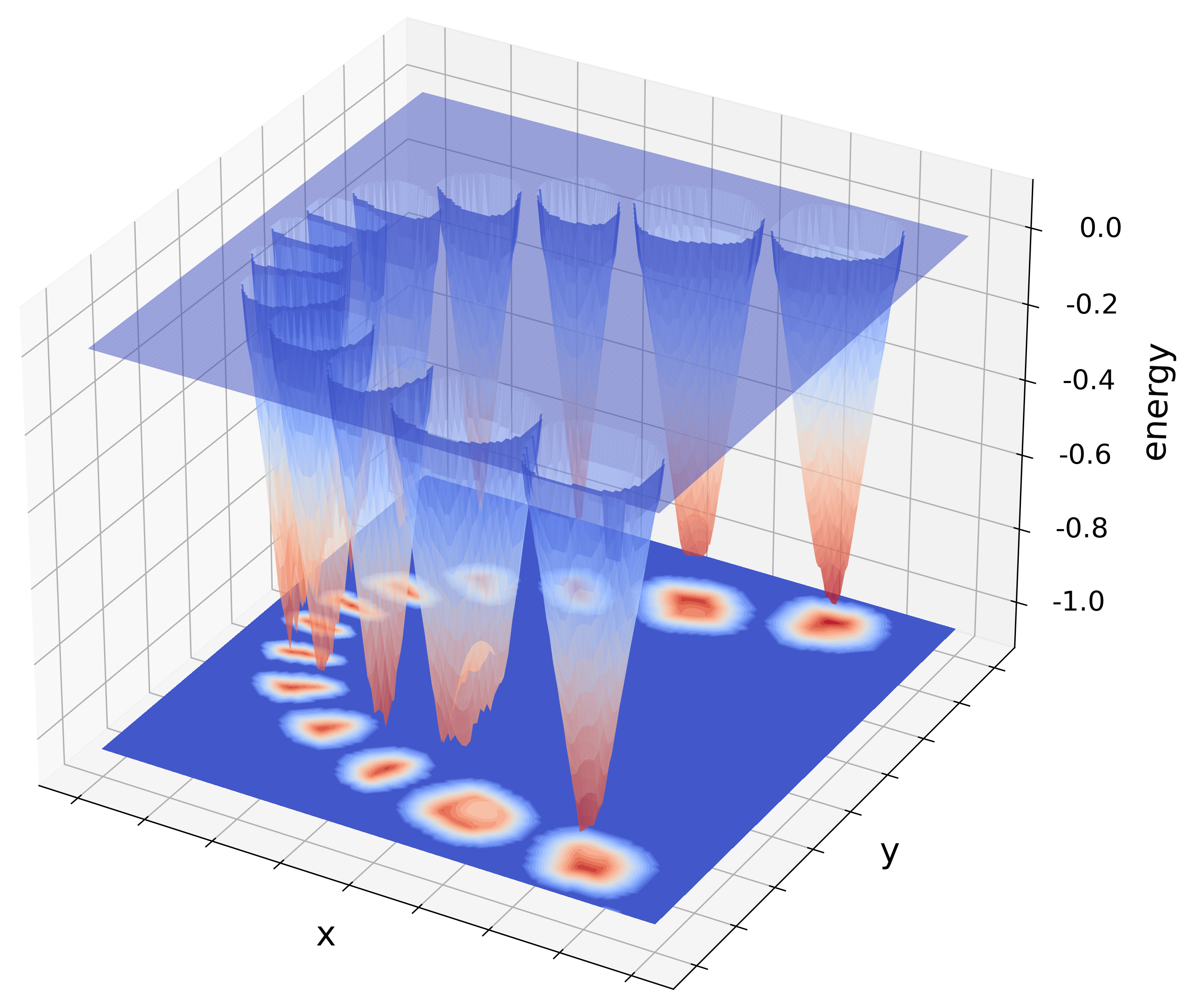}
\caption{Energy map in the axial ($xy$) plane. A representative slice from the 3D scan is shown, selected near the contact points where adjacent tooth crowns exhibit close anatomical proximity. This region, characterized by broad interproximal contact and parallel axial walls, presents a challenging condition for separating anatomically congruent and tangentially aligned teeth. The energy map highlights the model’s ability to resolve individual instances despite the absence of clear interproximal gaps or distinct morphological transitions between adjacent crowns.}
\label{fig:energy_map_EDT}
\end{figure}

\begin{figure}[!h]
\centering
\includegraphics[width=0.95\textwidth]{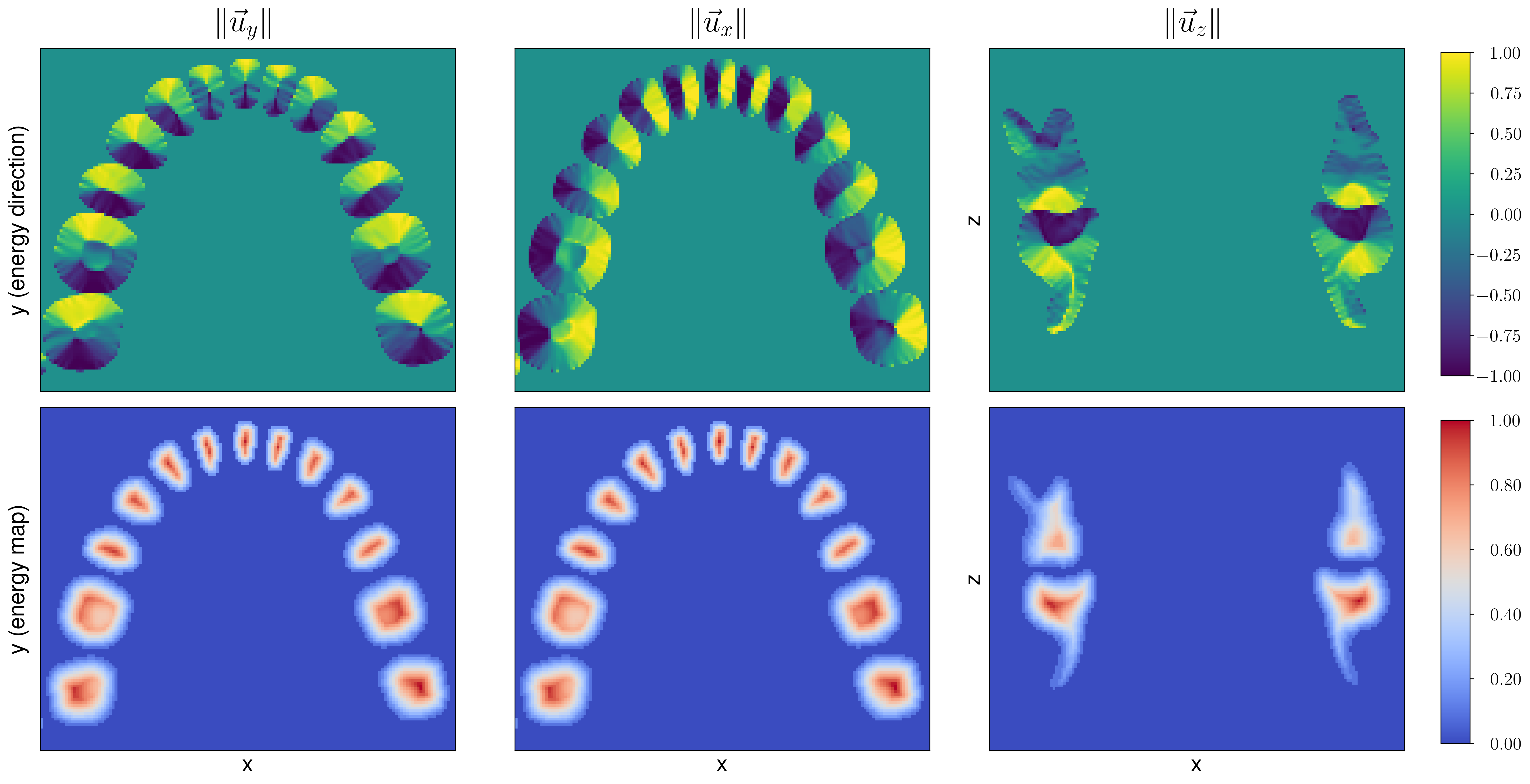}
\caption{Energy direction and energy map sections.}
\label{fig:EDT_angular_direction_scalar_field}
\end{figure}

To accurately delineate individual tooth instances in 3D CBCT scans, particularly in anatomically dense regions, we employ a deep watershed approach, which we adapt to 3D instance segmentation with particular focus on root apices. This method leverages a predicted scalar energy map to represent object centroids and a corresponding vector field to model directional gradients guiding instance separation. Figure~\ref{fig:energy_map_EDT} displays an axial slice ($xy$) of the energy map, selected near interproximal contact zones, where crown surfaces of adjacent teeth are in close tangential alignment. These regions are especially difficult to separate due to the absence of clear morphological discontinuities. Figure~\ref{fig:EDT_angular_direction_scalar_field} further illustrates the associated vector field components, showing how directionality in 3D space aids in disambiguating adjacent anatomical structures. This joint scalar-vector representation enables robust instance segmentation even under conditions of high anatomical congruence and tight contact.

Notably, Figure~\ref{fig:EDT_angular_direction_scalar_field} reveals rapid angular transitions in the vector field around anatomically intricate regions. This is especially pronounced at the root apices, where fine, tapering structures curve sharply and diverge from neighboring bone. The directional vector field exhibits high angular gradients in these areas, reflecting the need for fine-grained vector guidance to avoid root-level instance fusion. Similarly, the bite area, specifically in the $\| \overrightarrow{u_z} \|$ component representing superior-inferior direction, shows abrupt vector shifts due to the vertical overlap of opposing arches in occlusion. These steep gradients are critical for disambiguating closely apposed crowns from opposing jaws, enabling accurate watershed ridge formation along occlusal interfaces. This capacity to encode fast angular variation is essential for resolving complex 3D topologies in tight anatomical configurations.

\FloatBarrier
%_______________________________________________________________
\section{Qualitative Results}

\begin{figure}[!h]
\centering
\includegraphics[width=\textwidth]{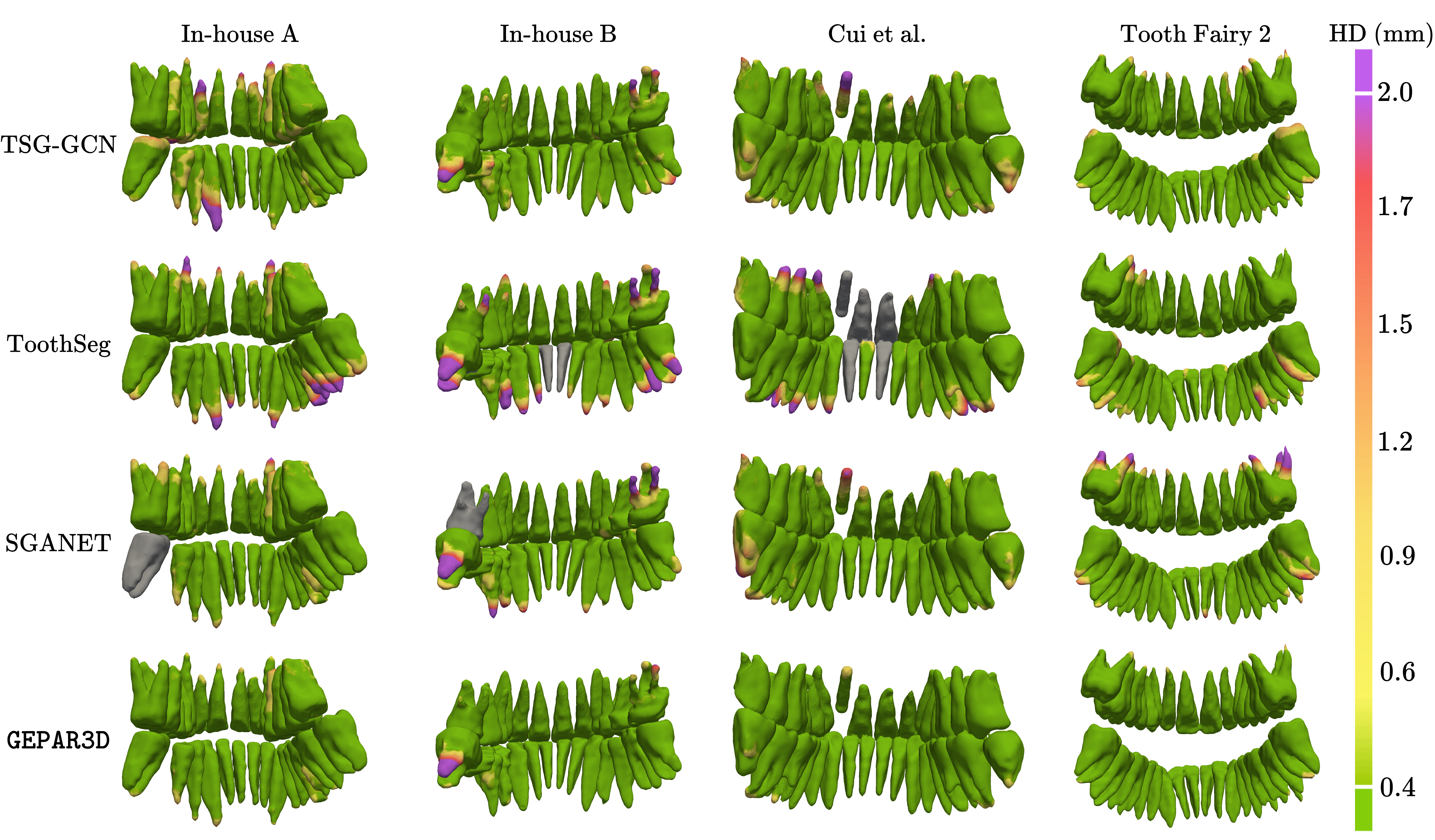}
\caption{More qualitative comparisons of GEPAR3D with Surface Hausdorff Distance heatmaps overlaid on GT labels (green = low, purple = high) highlight apex deviations. GEPAR3D shows superior root sensitivity versus tooth-specific baselines. Missing teeth are shown in gray.}
\label{fig:HD_error_heatmap}
\end{figure}

\begin{figure}[!h]
\centering
\includegraphics[width=\textwidth]{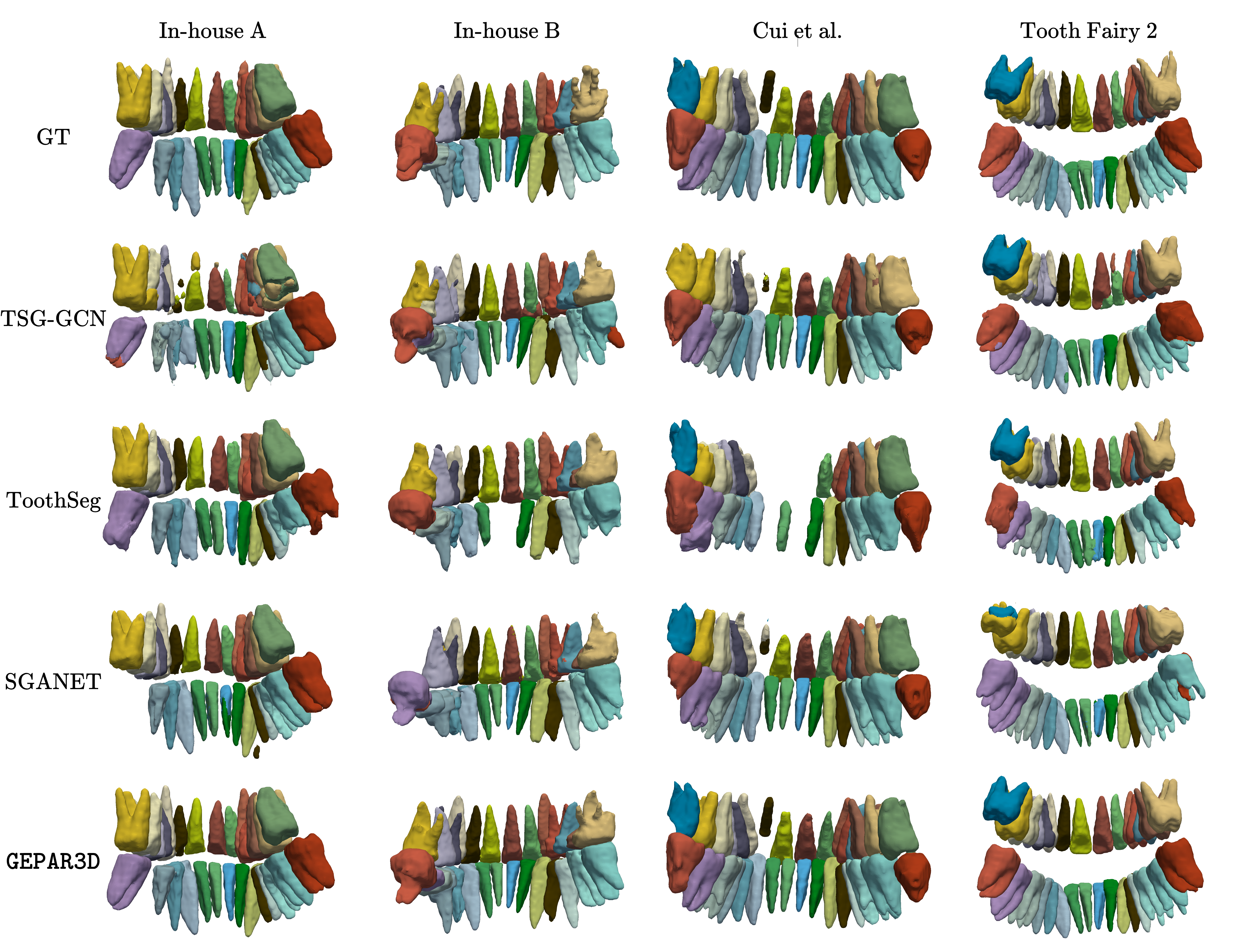}
\caption{Qualitative comparisons of segmentation on external test sets. We present scans corresponding to Fig. \ref{fig:HD_error_heatmap}. We present the raw 32-class segmentation results.}
\label{fig:qualitative_raw_gt}
\end{figure}

We present extended qualitative comparisons of GEPAR3D across all test datasets. Figure~\ref{fig:HD_error_heatmap} visualizes surface Hausdorff Distance (HD) heatmaps overlaid on ground truth labels, emphasizing segmentation errors near root apices, regions particularly prone to under-segmentation due to their fine geometry and low contrast, where other methods tend to over-smooth details. GEPAR3D demonstrates consistently improved sensitivity in these clinically critical areas compared to existing tooth-specific baselines. The raw segmentation outputs for the same samples are shown in Figure~\ref{fig:qualitative_raw_gt}, using a colormap consistent with Figure~\ref{fig:american_system} to depict individual tooth labels.

Unlike previous works that focused on challenges such as missing teeth or metal artifacts, we identify the precise delineation of root apices as an unresolved problem. Accurate root segmentation is essential for longitudinal assessments of root resorption, a condition where dental root structure is progressively lost. Since resorption is typically diagnosed by comparing serial CBCT scans, it requires a reliable baseline segmentation of the full root anatomy. No existing public CBCT dataset contains labeled resorbed cases, so in this work, we first establish apex segmentation performance on non-pathological datasets. Under-segmentation at baseline could falsely mask root shortening in follow-up. Future work will evaluate model performance in resorbed cases, as discussed in the main paper’s conclusions.

\end{document}